\documentclass{aa}  
\usepackage{natbib}
\bibpunct{(}{)}{;}{a}{}{,}
\usepackage{graphicx}
\usepackage[percent]{overpic}
\usepackage{xcolor}
\usepackage{epstopdf}
\usepackage{pdflscape}
\usepackage{array}
\usepackage{url}
\begin{document}

\authorrunning{Chatzistergos et al.}
\titlerunning{Ca~II~K emission vs. magnetic field}
   \title{Recovering the unsigned photospheric magnetic field from Ca~II~K observations}
\author{Theodosios Chatzistergos\inst{1, 2}, Ilaria Ermolli\inst{2}, Sami K. Solanki\inst{1,3}, Natalie A. Krivova\inst{1}, Fabrizio Giorgi\inst{2},  Kok~Leng~Yeo\inst{1}}
\offprints{Theodosios Chatzistergos  \email{chatzistergos@mps.mpg.de, theodosios.chatzistergos@inaf.it}}
\institute{Max Planck Institute for Solar System Research, Justus-von-Liebig-weg 3,	37077 G\"{o}ttingen, Germany \and INAF Osservatorio Astronomico di Roma, Via Frascati 33, 00078 Monte Porzio Catone, Italy \and School of Space Research, Kyung Hee University, Yongin, Gyeonggi 446-701, Republic of Korea}
\date{}

   \abstract
{A number of studies have aimed at defining the exact form of the relation between magnetic field strength and Ca II H and K core brightness.  All previous studies have, however, been restricted to isolated regions on the solar disc or to a limited set of observations.}
{We reassess the relationship between the photospheric magnetic field strength and the Ca II K intensity for a variety of surface features as a function of the position on the disc and the solar activity level. This relationship can be used to recover the unsigned photospheric magnetic field from images recorded in the core of Ca II K line.} 
{We have analysed 131 pairs of high-quality, full-disc, near-co-temporal observations from the Helioseismic and Magnetic Imager (SDO/HMI) and Precision Solar Photometric Telescope (Rome/PSPT) spanning half a solar cycle. To analytically describe the observationally-determined relation, we considered three different functions: a power law  with an offset, a logarithmic function, and a power law function of the logarithm of the magnetic flux density. We used the obtained relations to reconstruct maps of the line-of-sight component of the unsigned magnetic field (unsigned magnetograms) from Ca II K observations, which were then compared to the original magnetograms.} 
{We find that both power-law functions represent the data well, while the logarithmic function is good only for quiet periods. We see no significant variation over the solar cycle or over the disc in the derived fit parameters, independently of the function used. We find that errors in the independent variable, usually not accounted for, introduce attenuation bias. To address this, we binned the data with respect to the magnetic field strength and Ca II K contrast separately and derived the relation for the bisector of the two binned curves. The reconstructed unsigned magnetograms show good agreement with the original ones. RMS differences are less than 90 G. The results were unaffected by the stray-light correction of the SDO/HMI and Rome/PSPT data. } 
{Our results imply that Ca~II~K observations, accurately processed and calibrated, can be used to reconstruct unsigned magnetograms by using the relations derived in our study.} 


\maketitle

\section{Introduction}

\begin{figure}[t!]
	\centering
	\includegraphics[width=1\linewidth]{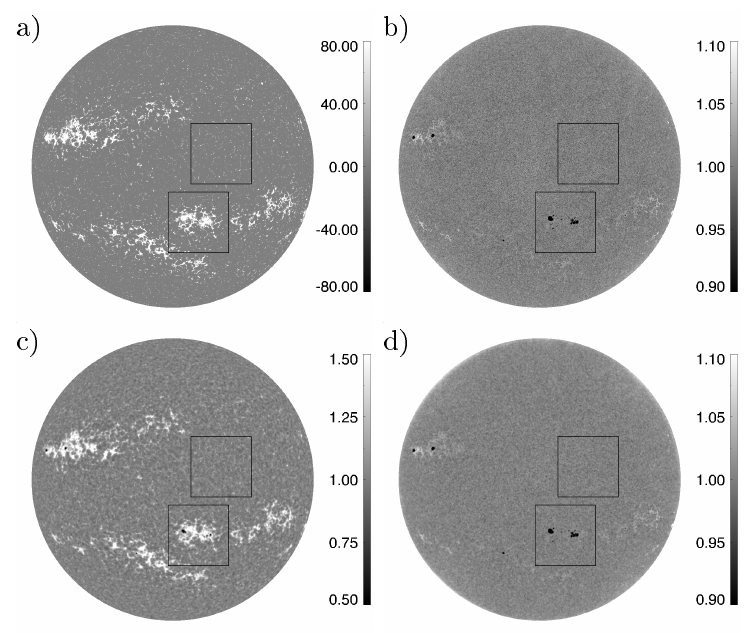}
	\caption{Examples of the observations analysed in this study taken on 01/04/2011 at 09:06:00 UT: (a)) SDO/HMI unsigned $B_\mathrm{LOS}$ magnetogram, (b)) SDO/HMI continuum contrast image (i.e. compensated for intensity CLV),  (c)) Rome/PSPT Ca~II~K and (d)) red continuum contrast images. The grey-scale bars on the right-hand side of each panel show magnetic signal in G and contrast, respectively. The squares indicate the insets shown in Fig. \ref{fig:networkzoom}.}
	\label{fig:psptcaexamples}
\end{figure}

\newcounter{tableidbvscak}

\begin{table*}
	\caption{Previously published results on the relation between magnetic flux and Ca II intensity along with our results.}
	\label{tab:previousstudies}
	\centering
	\tiny
	{\setlength{\extrarowheight}{3pt}%
		\begin{tabular}{lccccccccc}
			\hline\hline
			Ref. & Line & Bandwidth &N & Period	    		   & Type	  & Location & Region size      & Pixel scale & Relation \\
			  	 &      & [\AA{}]   &  &     		   		   &  	      &          &                  &  [$''$]      &   \\
			\hline
			\addtocounter{tableidbvscak}{1}\thetableidbvscak	& K & 1.1 &7    & 21--27/10/1968		   & AR & disc & $2.4''\times 2.4'' $, $1''\times 1'' $  & $2.4$, $1$  & polynomial\\ 
			\addtocounter{tableidbvscak}{1}\thetableidbvscak	& K & 1.1 & 1    & 09/1968     		   & QS & centre & $2.4''\times 2.4'' $  & $2.4$  & linear\\  
			\addtocounter{tableidbvscak}{1}\thetableidbvscak	& K	& 3.2 & 1    & 15/10/1987 		   & AR & centre & $256''\times 360''$ 	   &$4$ & binning\tablefootmark{a}\\ 
			\addtocounter{tableidbvscak}{1}\thetableidbvscak	& K	& 0.1 & 1    & 22/10/1985 		   & AR & centre & $390''\times 540''$ 	   &$2.4$ & power law\\ 
			\addtocounter{tableidbvscak}{1}\thetableidbvscak	& K	& 0.1 & 1    & 21/12/1994 		   & AR & centre & $60^\circ\times 40^\circ$ 	   &$4$ & power law\\ 
			\addtocounter{tableidbvscak}{1}\thetableidbvscak	& K	& 0.3 & 2    & 13/10/1996, 13/04/1997 & QS & centre & $\sim 170''\times 160''$ & $2$& linear\\ 
			\addtocounter{tableidbvscak}{1}\thetableidbvscak	& K	& 0.5 & 8    & 16/01/1992--08/07/1993   & QS+AR	& \multicolumn{2}{c}{full-disc} & $2$	   & power law\\ 
			\thetableidbvscak	& K	& 3 & 7    & 16/01/1992--08/07/1993   & QS+AR	& \multicolumn{2}{c}{full-disc} & $4$	   & power law\\ 
			\thetableidbvscak	& K	& 10 & 4   & 03/06/1993--08/07/1993   & QS+AR	& \multicolumn{2}{c}{full-disc} & $2$	   & power law\\ 
			\addtocounter{tableidbvscak}{1}\thetableidbvscak	& K & 1.2 & 2    & 2000				   & QS+AR & centre &$\sim 470''\times 470''$ & $2$& binning\tablefootmark{a}\\ 
			\addtocounter{tableidbvscak}{1}\thetableidbvscak	& K	& 3 & -    & 2005 6 days 		   & AR & centre & $\sim 810''\times 810''$ & $2$& power law\\ 
			\addtocounter{tableidbvscak}{1}\thetableidbvscak 	& K & 3 & 60   &  28/05/1999--31/07/1999 		   & QS+AR & \multicolumn{2}{c}{full-disc} 	   & $1$& power law\\ 
			\addtocounter{tableidbvscak}{1}\thetableidbvscak 	& H & 0.5 & 13   & 27/09/2004  		   & QS & centre & $25.5''\times25.5''$ 	   & $1$& power law\\ 
			\addtocounter{tableidbvscak}{1}\thetableidbvscak	& K & 0.6 & 1    & 18/05/2004  		   & QS & centre & $300''\times300''$ 	   & $4$& power law\\ 
			\addtocounter{tableidbvscak}{1}\thetableidbvscak    & IR& 16.1 & 2    &  20/04/2015, 13/05/2015    & QS+AR & \multicolumn{2}{c}{full-disc}  & $1$& total flux $\varpropto$ plage area\\ 
			\thetableidbvscak   & K &  0.35 & -   &  1973--1985 		   & QS+AR &  \multicolumn{2}{c}{carrington maps}  & & total flux $\varpropto$ plage area\\  
			\addtocounter{tableidbvscak}{1}\thetableidbvscak    & H & 1.8 & 40   &	09/06/2009  	   & QS & centre & $50''\times50''$  & $0.1$ &logarithmic\\ 
			\addtocounter{tableidbvscak}{1}\thetableidbvscak    & H & 1.1 & 28   &	12/06/2013  	   & QS & AR & $15''\times38''$  & $0.02$ &logarithmic\\ 
			\addtocounter{tableidbvscak}{1}\thetableidbvscak & K & 2.5 & 131 & 18/05/2010--29/08/2016 & QS+AR&\multicolumn{2}{c}{full-disc} & $2$ & logarithmic power law\\
			\hline\hline	
	\end{tabular}}
	\tablefoot{Columns are: reference, spectral line, bandwidth, number and period of observations, type, location, and dimensions of the analysed region, the pixel scale, and the type of relation derived. Dashes denote missing information. \tablefoottext{a}{These studies did not derive the functional form of the relation, they simply binned  the available datapoints with respect to the magnetic field strength. We note, however, that the results they presented are approximately consistent with a power law function.}}
	\tablebib{\addtocounter{tableidbvscak}{-\thetableidbvscak}
		(\addtocounter{tableidbvscak}{1}\thetableidbvscak)~\citet{frazier_multi-channel_1971}; (\addtocounter{tableidbvscak}{1}\thetableidbvscak)~\citet{skumanich_statistical_1975};  (\addtocounter{tableidbvscak}{1}\thetableidbvscak) \citet{wang_relationship_1988}; (\addtocounter{tableidbvscak}{1}\thetableidbvscak) \citet{schrijver_relations_1989};  (\addtocounter{tableidbvscak}{1}\thetableidbvscak) \citet{schrijver_dynamics_1996}; (\addtocounter{tableidbvscak}{1}\thetableidbvscak) \citet{nindos_relation_1998}; (\addtocounter{tableidbvscak}{1}\thetableidbvscak) \citet{harvey_magnetic_1999}; 
		(\addtocounter{tableidbvscak}{1}\thetableidbvscak) \citet{rast_supergranulation:_2003}; (\addtocounter{tableidbvscak}{1}\thetableidbvscak) \citet{ortiz_how_2005}; (\addtocounter{tableidbvscak}{1}\thetableidbvscak) \citet{vogler_solar_2005}; (\addtocounter{tableidbvscak}{1}\thetableidbvscak) \citet{rezaei_relation_2007}; (\addtocounter{tableidbvscak}{1}\thetableidbvscak) \citet{loukitcheva_relationship_2009}; (\addtocounter{tableidbvscak}{1}\thetableidbvscak) \citet{pevtsov_reconstructing_2016}; (\addtocounter{tableidbvscak}{1}\thetableidbvscak) \citet{kahil_brightness_2017}; (\addtocounter{tableidbvscak}{1}\thetableidbvscak) \citet{kahil_intensity_2019}; (\addtocounter{tableidbvscak}{1}\thetableidbvscak) This work, see Sect. \ref{sec:bvscakresults}.
	}
\end{table*}

\citet{babcock_suns_1955} noticed a  ``one-to-one correspondence'' between bright regions in Mt Wilson Ca~II~K spectroheliograms and magnetic regions in magnetograms. This reported association, which was promptly confirmed by  \citet{howard_observations_1959} and \citet{leighton_observations_1959}, has
initiated numerous studies of solar and stellar Ca II data. 
Since then, considerable efforts have been devoted to understand the relation between the magnetic field strength and the Ca~II~K intensity for different solar magnetic regions on the Sun \citep[e.g.][]{frazier_multi-channel_1971,skumanich_statistical_1975,schrijver_relations_1989,nindos_relation_1998,harvey_magnetic_1999,vogler_solar_2005,rast_scales_2003,ortiz_how_2005,rezaei_relation_2007,loukitcheva_relationship_2009,pevtsov_reconstructing_2016,kahil_brightness_2017,kahil_intensity_2019}. 
Table \ref{tab:previousstudies} summarises the main features and results of the earlier studies compared with the results of this one. All  previous works were based on analysis of small data samples \citep[with the possible exception of][]{vogler_solar_2005}, mainly considering regions at the disc centre, and using data with a spatial resolution lower than 
$\approx~2\arcsec$. Most of earlier studies reported that the link between the magnetic field strength and Ca II K intensity is best described by a power law function with exponent in the range 0.3-0.6. However, \cite{skumanich_statistical_1975} and \cite{nindos_relation_1998} found that their data were best represented by a linear relation, while \cite{kahil_brightness_2017,kahil_intensity_2019} found a logarithmic function to fit best their data. It is worth noting that the latter authors analysed Ca II H observations taken with the Sunrise balloon-borne telescope \citep{solanki_sunrise:_2010,solanki_second_2017,barthol_sunrise_2011}, which have a higher  spatial resolution than in previous studies. These (and other similar) studies are discussed in detail in Sec. \ref{sec:somparisontoothers}.

Major efforts have also been invested to measure the disc-integrated Ca II H and K emission of many other stars. Such measurements have been regularly carried out e.g. within the synoptic ground-based programs at Mt Wilson \citep[1966--2003, ][]{wilson_chromospheric_1978,duncan_ca_1991,baliunas_chromospheric_1995} and Lowell Observatories \citep[1994--present, ][]{hall_activity_2007}, as well as by the space-born photometer on-board the CoRoT mission \citep{michel_first_2008,auvergne_corot_2009,gondoin_corot_2012}. 
Ca II H $\&$ K emission is an
indicator of the strength of, and the area covered by,  magnetic fields on the Sun \citep{leighton_observations_1959}. Since the Ca II H and K variations due to magnetic regions are of the order of a few tens of percent, they can be easily detected for many active stars. Hence, the Ca II H and K measurements have been used 
to trace long-term changes in surface activity  of stars caused by e.g. the activity cycle, rotation, and convection \citep[e.g.][etc.]{sheeley_average_1967,white_solar_1978,keil_variations_1984,baliunas_time-series_1985}. These studies have led to an improved knowledge of stellar rotation and activity, and of the degree to which the Sun  and other stars share similar dynamical properties  \citep[for reviews, see e.g.][]{lockwood_patterns_2007,lockwood_decadal_2013,hall_stellar_2008,reiners_observations_2012}. It is worth noting that stellar Ca II observations are per force integrated over the whole stellar disc. However, except for the studies of \citet{harvey_magnetic_1999,vogler_solar_2005}, and \cite{pevtsov_reconstructing_2016}, restricted to a few images, no other previous investigation has determined the relation between Ca II brightness and magnetic field strength covering the full solar disc.

Furthermore, many studies require long data sets of the solar surface magnetic field, e.g. to derive information on the  structure, activity, and variability of the Sun, or for related applications such as Earth's climate response to solar irradiance variability.  Regular magnetograms are, however, available only for the last four solar cycles, while synoptic Ca II K solar observations have been carried out for more than 120 years \citep{chatzistergos_analysis_2017,chatzistergos_analysis_2019}. In recent years, following the availability of a number of digitized series of historical Ca~II~K observations, attempts have been made to reconstruct magnetograms from Ca~II~K observations, based on the relation between the Ca~II~K intensity and the magnetic field strength. In particular, \citet{pevtsov_reconstructing_2016} reconstructed magnetograms from Ca~II~K synoptic charts made from Mt Wilson observatory images. For their reconstruction they used sunspot records to get information about the polarity and assigned each plage area with a single magnetic field strength value based on the area of the plage. The areas and locations of plage regions were derived from photometrically uncalibrated Ca~II~K images. 
Besides that, \cite{sheeley_carrington_2011} and \cite{chatterjee_butterfly_2016} constructed Carrington maps with Ca~II~K images from the Mt Wilson and Kodaikanal observatories, respectively. These maps can be used to trace the evolution of the plage regions. However they provide Ca~II~K contrast and need to be converted into magnetic field strength for any application based on magnetic field measurements.

In this paper, we study the relationship between the magnetic field strength and the Ca~II~K intensity using data from two archives of high-quality full-disc solar observations. 
We use significantly more data of higher quality than in previous studies, which allows a more detailed and accurate assessment of this relationship over the whole disc and at different levels of solar activity during cycle 24. 
We test the accuracy of our results by applying the derived relationship to reconstruct unsigned magnetograms and then comparing them with the actual ones.

This paper is organised as follows. Section \ref{sec:datamethodsmain} describes the  data and methods employed  for our analysis. In Section \ref{sec:bvscakresults}  we study the magnetic field strength and the Ca~II~K excess intensity. In section \ref{sec:reconstructing_magnetograms} we use our results to reconstruct magnetograms from the Ca~II~K images and to test the accuracy of our method. 
Finally we draw our conclusions in Section \ref{sec:bvscaksummary}.

\begin{figure*}
	\centering
	\includegraphics[width=1\linewidth]{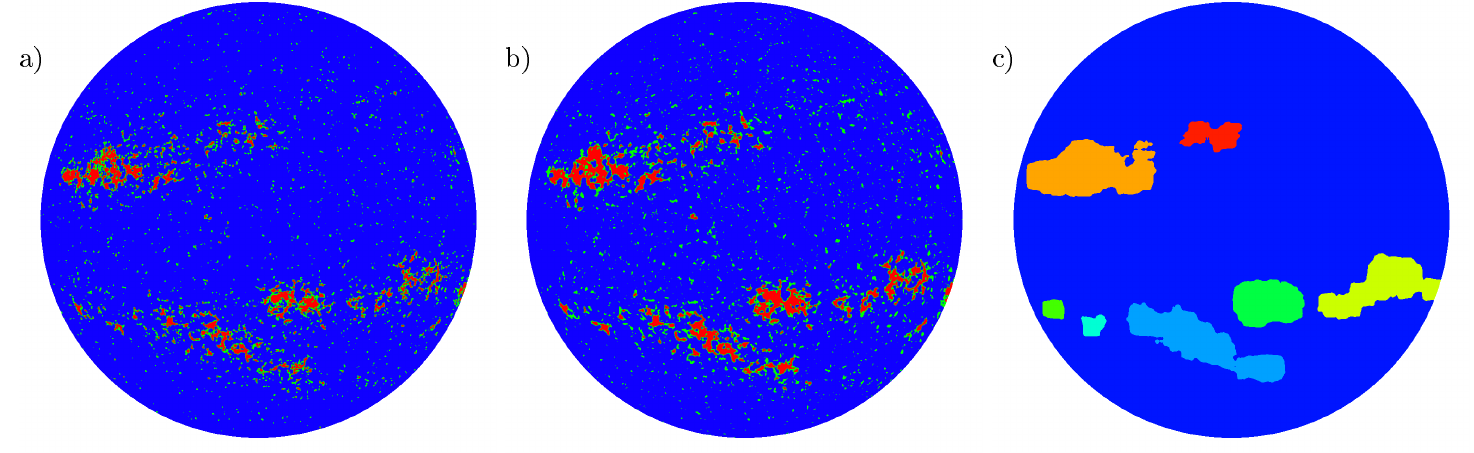}
	\caption{Segmentation masks of bright magnetic features derived from the observations shown in Fig. \ref{fig:psptcaexamples} by applying the 2 methods described in Sect. \ref{sec:segmentation}. (a)) Mask of magnetogram and (b))  Ca~II~K image derived with Method 1, showing plage (red), network (green), and QS (blue). (c)) Mask of magnetogram derived with Method 2, showing individual activity clusters with different colours (the QS is in dark blue). The masks are shown prior to the exclusion of the sunspot regions.}
	\label{fig:segmentationmasks}
\end{figure*}

\section{Data and methods}
\label{sec:datamethodsmain}
\subsection{Data}
\label{sec:datamethods}

We analysed full-disc photospheric longitudinal  magnetograms  and continuum intensity images from the space-borne Helioseismic and Magnetic Imager \citep[HMI, ][]{scherrer_helioseismic_2012,schou_design_2012} aboard the Solar Dynamics Observatory  \citep[SDO, ][]{pesnell_solar_2012}, and full-disc filtergrams taken at the Ca~II~K line and red continuum from the Precision Solar Photometric Telescope at the Rome Observatory \citep[Rome/PSPT, ][]{ermolli_prototype_1998,ermolli_photometric_2007}.
Figure \ref{fig:psptcaexamples} shows examples of the analysed SDO/HMI and Rome/PSPT images. 

Rome/PSPT, in operation since 1996, is a 15 cm telescope designed  for photometric solar observations characterized
by  0.1\% pixel-to-pixel 
relative photometric precision
\citep{coulter_rise/pspt_1994}. 
The images\footnote{Available at \url{http://www.oa-roma.inaf.it/fisica-solare/}} were acquired  with narrow-band  
interference filters, by single exposure of a 2048$\times 2048$ CCD array. The filters employed for the observations analysed here are  centred at the Ca~II~K line core (393.3 nm) with bandwidth of 0.25 nm, and in the red continuum at 607.2 nm with bandwidth of 0.5 nm.  The Ca II K and red continuum images were taken within 3 minutes from each other. At the acquisition, the data were reduced to a pixel scale of $2''$ to account for typical  conditions of local seeing. 
Standard instrumental calibration has been applied to the data \citep[][]{ermolli_prototype_1998,ermolli_radiative_2010}.

SDO/HMI, in operation since April 2010, takes full-disc 4096$\times$4096 pixel filtergrams at six wavelength positions across the Fe I 617.3 nm line at 1.875 s intervals. The filtergrams are combined to form simultaneous continuum intensity images and longitudinal magnetograms with a pixel scale of $0.505''$ and 45 s cadence. 
For each Rome/PSPT image pair, we took the 360 s average of the SDO/HMI images and magnetograms taken close in time (on average less than 2 minutes apart and no more than 8 minutes).
The averaging was done to suppress intensity and magnetogram signal fluctuations from noise and $p$-mode oscillations. 

For our analysis, we have selected data with the highest spatial resolution (for Rome/PSPT), closest time between SDO/HMI and Rome/PSPT observations, and highest signal-to-noise ratio.
We avoided winter periods and kept observations mostly during summer months, when the seeing-induced degradation in Rome/PSPT data is lower.
Our data sample consists of 131 sets of near-simultaneous observations. 
These observations cover the period between 18/05/2010 and 29/08/2016.

We have ignored the pixels in SDO/HMI magnetograms with flux density below 20 G. The value of 20 G corresponds roughly to 3 times of the noise level as evaluated by \cite{yeo_intensity_2013,yeo_reconstruction_2014}. Since the magnetic flux tubes making up network and plage tend towards an orientation normal to the surface, while magnetograms measure the line-of-sight (LOS, hereafter) component of it ($B_{\mathrm{LOS}}$), we divided the pixel signal by the corresponding $\mu$ (cosine of the heliocentric azimuthal angle) to get the intrinsic magnetic field strength. 
We also removed the polarity information from the SDO/HMI data, and only considered the absolute value of the magnetic flux density, $|B_{\mathrm{LOS}}|/\mu$ (i.e. the magnetic field strength averaged over the effective pixel). 

The Rome/PSPT images were first rescaled to match the size of SDO/HMI so that we could align both observations with highest accuracy. The Rome/PSPT images  were then rotated and aligned to the SDO/HMI observations, by applying compensations for ephemeris.
All observations were then re-scaled to the original dimensions of Rome/PSPT.
To further reduce effects due to seeing, we also reduced the resolution of the SDO/HMI data to that of the Rome/PSPT by smoothing them with a low-pass filter with a $2\times2$ pixel running window width. In the following, we refer to the SDO/HMI data so obtained as SDO/HMI degraded magnetograms.

For each analysed intensity image (Rome/PSPT and SDO/HMI) we removed the limb darkening and obtained a contrast map. In particular, for each image pixel $i$, we defined its contrast $C_i$ as $C_i = I_i/I^{\mathrm{QS}}_{i}$, where $I_i$ is the measured intensity of pixel $i$, and $I^{\mathrm{QS}}_{i}$ is the intensity of the quiet Sun (QS, hereafter) at the same position. The latter was derived with  the iterative procedure described by \cite{chatzistergos_analysis_2018}, which returns contrast images with an average error in contrast values lower than 0.6\% \citep[see][for more details]{chatzistergos_analysis_2018}. 

Since our aim here was a study of the relation between the magnetic field strength and Ca II K brightness in bright magnetic regions, we masked out sunspots in the magnetograms and in Ca~II~K observations. Sunspots were identified in SDO/HMI continuum intensity as regions having intensity contrast lower than 0.89 \citep[following][]{yeo_intensity_2013} and in Rome/PSPT red continuum images lower than 0.95. The above thresholds were derived as the average value of $\bar{C}-3\sigma$, where $\bar{C}$ is the average value of contrast over the disc and $\sigma$ is the standard deviation of contrast values, from all Rome/PSPT red continuum and SDO/HMI continuum images separately. 
The plage regions immediately surrounding sunspots were excluded as well, as they could be affected by stray-light and by extended low-lying sunspot canopies \citep[e.g.][]{giovanelli_three-dimensional_1982,solanki_infrared_1994,solanki_expansion_1999}, as was shown by \citet{yeo_intensity_2013}. This was done by expanding the sunspot regions with a varying size kernel, corresponding to $10\times10$ and $30\times30$ pixel$^2$ at disc centre and limb, respectively. 
The excluded regions have areas on average 0.001 in fraction of the disc, while the maximum value is 0.005. These regions amount on average to $13\pm9$\% of the total flux in the original magnetograms, which appears to be roughly constant in time for the analysed data.

\begin{figure*}[t!]
	\centering
	\includegraphics[width=1.0\linewidth]{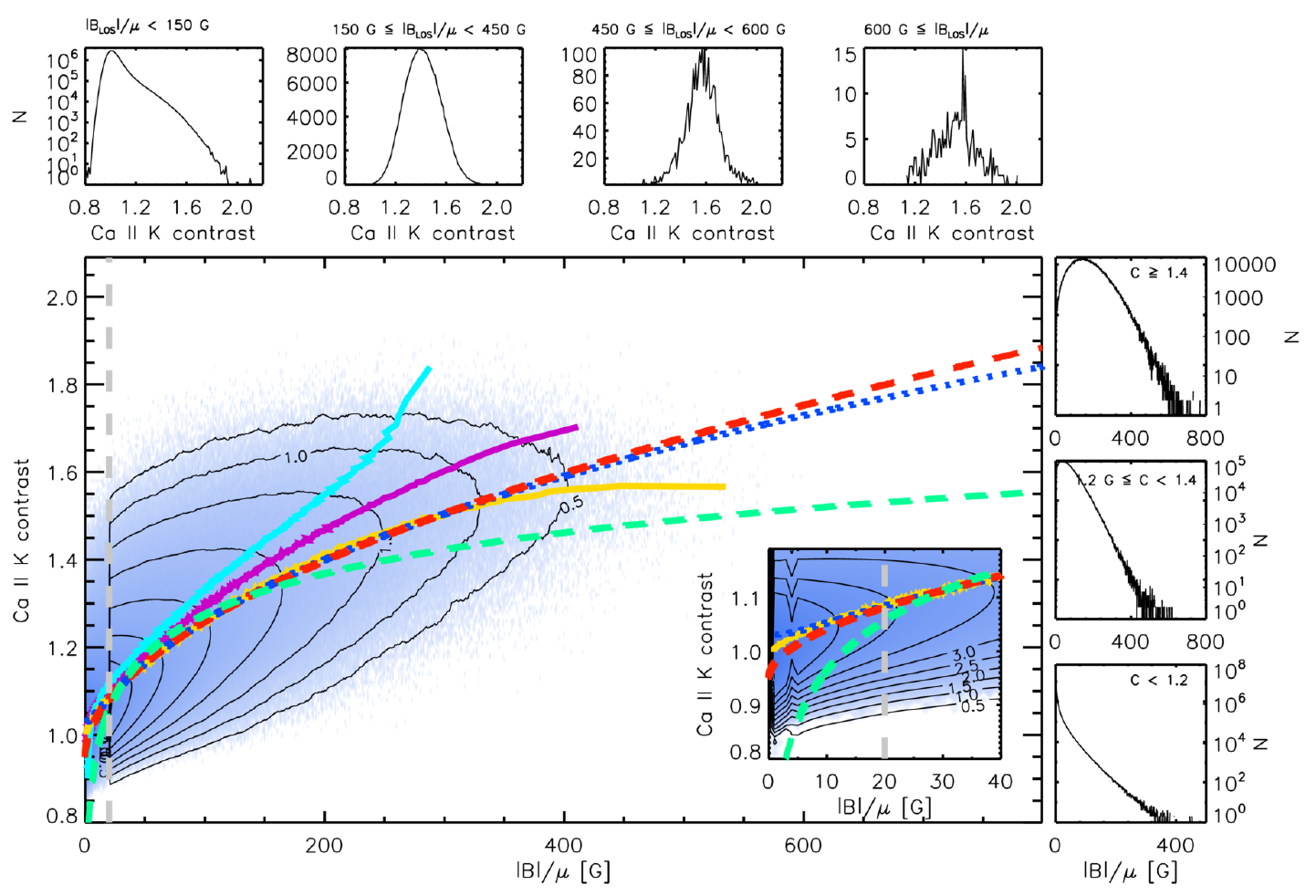}
	\caption{Ca~II~K contrast plotted against the unsigned LOS magnetic flux density divided by $\mu$ ($|B_{\mathrm{LOS}}|/\mu$) for all pixel pairs (excluding sunspots) in all available images.  The pixels are colour-coded denoting the logarithm of the number density within bins of 1 G and 0.01 in contrast. The contour lines give the logarithm of the pixel number density in intervals of 0.5.
		Curves show 5000-point running means (over $|B_{\mathrm{LOS}}|/\mu$ in yellow, over contrast in light blue, and their bisector in purple), as well as power law function (PF, red), power law function of $\log{(|B_{\mathrm{LOS}}|/\mu)}$ (PFL, blue) and logarithmic function (LFL, green) fits on the binned curve over $|B_{\mathrm{LOS}}|/\mu$ (yellow curve). The vertical grey dashed line denotes the 20 G threshold in $|B_{\mathrm{LOS}}|/\mu$. A magnified section for low $|B_{\mathrm{LOS}}|/\mu$ is shown at the lower right corner of the figure to illustrate the differences of the different fits over that region. Also shown are histograms of $|B_{\mathrm{LOS}}|/\mu$ within 3 ranges of contrast values (right) and histograms of Ca~II~K contrast for 4 ranges of $|B_{\mathrm{LOS}}|/\mu$ (top). The ranges used for the histograms are shown on the upper part of the corresponding sub-plot.}
	\label{fig:fitexample}
\end{figure*}

\subsection{Stray-light removal}
\label{sec:straylight}
To investigate whether our results depend on the removal of stray-light from the analysed images, we restored 51 pairs of the SDO/HMI and Rome/PSPT images following \citet{yeo_point_2014} and \citet{criscuoli_stray-light_2008}, respectively. We have also analysed a sample of 10 SDO/HMI magnetograms from our dataset that were restored with the method employed by \cite{criscuoli_photometric_2017}.
Employment of different methods helps us to assess the potential errors in the relation between the Ca II K contrast and the magnetic field strength due to the stray-light degradation.

For the SDO/HMI observations, the point-spread function (PSF, hereafter) of the instrument was deconvolved from Stokes I and V observables that were then used to produce the stray-light corrected magnetograms. 
The PSF derived by \citet{yeo_point_2014} has the form of the sum of five Gaussian functions. The PSF parameters were determined from the Venus transit data by performing a fit over the shaded areas. 

The PSF applied by \cite{criscuoli_photometric_2017} instead has the form of an Airy function convolved with a Lorentzian. The parameters of the PSF were derived by using pre-launch testing data as well as post-launch off-limb data taken during a partial lunar eclipse and the transit of Venus. According to \cite{criscuoli_photometric_2017}, the PSF employed by \citet{yeo_point_2014} does not account for large-angle or long-distance scattering, thus affecting results from analyses of data concerning large spatial scales on the solar disc such as in the present study. 

The Rome/PSPT data were deconvolved by using analytical functions defined from modelling the centre-to-limb variation of intensity in the data and instrumental PSF \citep{criscuoli_stray-light_2008}. The PSF here is modelled as the sum of three Gaussian and one Lorentzian functions, following \citet{walton_restoration_1999}.

\subsection{Segmentation}
\label{sec:segmentation}
For our analysis we selected pixels that correspond to magnetic regions in magnetograms and bright regions in Ca~II~K images.  
We identified features of interest with two methods.

\textit{Method 1.} We distinguished between two different types of bright magnetic features: plage and the network. They are differentiated with single contrast and  $|B_{\mathrm{LOS}}|/\mu$ thresholds in  Ca~II~K and magnetograms, respectively.
The thresholds are 20 G$\leq|B_{\mathrm{LOS}}|/\mu<60$ G and $1.12\leq C<1.21$ for network and  $|B_{\mathrm{LOS}}|/\mu\geq60$ G and $C\geq 1.21$ for plage. 
The thresholds given above for plage in the magnetograms, as well as for the network in Ca~II~K images, were acquired by minimising the differences between the average disc fractions calculated in the magnetograms and the Ca~II~K images.

\textit{Method 2.} 
We used this method to isolate individual activity clusters which may be composed of multiple close or overlapping active regions (ARs, hereafter). In this way we can study how the relation between the magnetic field strength and the Ca~II~K contrast varies among features of different sizes and locations on the disc. We applied a low-pass filter with a 50-pixel window width to the degraded magnetograms and a constant threshold of $|B_{\mathrm{LOS}}|/\mu=15$ G to isolate individual magnetic regions. Contiguous pixels were grouped together, and all isolated regions were considered as separate clusters. We also applied a size threshold of 50 pixels to the clusters. Pixels not assigned to any cluster were categorised as QS, though they include the network as well. This method is similar to that used by \cite{harvey_magnetic_1999}.

In our analysis we excluded all pixels with $\mu < 0.14$ (outermost 1\% of the solar radius) to restrict errors due to projection effects. Finally, the sunspot regions were also excluded from all masks as described in Sect. \ref{sec:datamethods}.

Figures \ref{fig:segmentationmasks}a) and \ref{fig:segmentationmasks}b) show the masks derived from SDO/HMI magnetogram and the Rome/PSPT Ca~II~K image with Method 1 on the images shown in Figs. \ref{fig:psptcaexamples}a) and \ref{fig:psptcaexamples}c), respectively. 
Plage regions are shown in red, network in green and QS in blue.
Figure \ref{fig:segmentationmasks}c) shows the mask derived with Method 2 on the Ca II K image shown in Fig. \ref{fig:psptcaexamples}a). The different features are shown with different colours, while the QS is shown in dark blue. 
\begin{figure}[t!]
	\centering
	\includegraphics[width=1\linewidth]{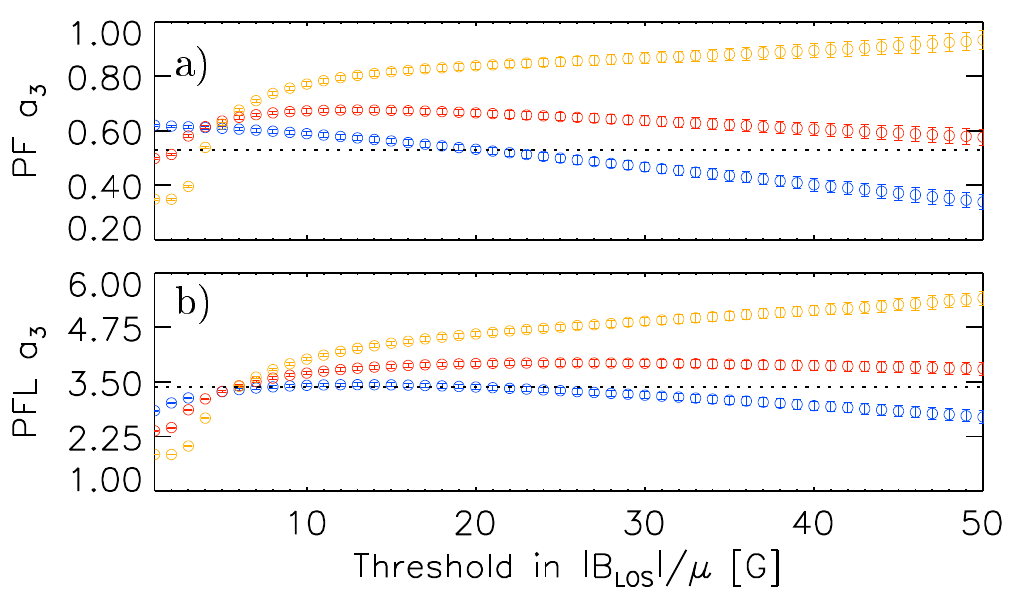}
	\caption{Parameters of the fits (Eq. \ref{eq:logfit1}) as a function of the threshold in $|B_{\mathrm{LOS}}|/\mu$ for PF (a)) and PFL (b)) fits. The fits are performed on the curves derived by binning over $|B_{\mathrm{LOS}}|/\mu$ (blue), contrast values (yellow), and the bisector (red). The dotted line in each panel is the best fit parameter derived with the threshold of $|B_{\mathrm{LOS}}|/\mu=20$ G for the blue points.}
	\label{fig:exponents_wholediscfeatures_lowthreshtest}
\end{figure}

\section{Results}
\label{sec:bvscakresults}

\subsection{Pixel-by-pixel relationship}
\label{sec:scatterpots}

\begin{table*}
	\caption{Results of fitting.} 
		\label{tab:functionparameters}
	\centering
	{\setlength{\extrarowheight}{3pt}%
		\begin{tabular}{lcccccc}
			\hline\hline
			Function & $x$ & Binning & $a_1$ & $a_2$ & $a_3$ & $\chi^2$ \\
			\hline
			PF & $|B_{\mathrm{LOS}}|/\mu$ 			&$|B_{\mathrm{LOS}}|/\mu$& $0.950\pm0.006$ & $0.027\pm0.002$ & $0.53\pm0.01$ & 0.16\\ 
			PF & $|B_{\mathrm{LOS}}|/\mu$ 			&$C$& $1.043\pm0.003$ & $0.006\pm0.001$ & $0.84\pm0.01$ & 0.09\\ 
			PF & $|B_{\mathrm{LOS}}|/\mu$ 			&bisector& $1.004\pm0.004$ & $0.014\pm0.001$ & $0.67\pm0.01$ & 0.03\\ 
			PFL & $\log{(|B_{\mathrm{LOS}}|/\mu)}$  &$|B_{\mathrm{LOS}}|/\mu$& $1.031\pm0.003$ & $0.022\pm0.001$ & $3.38\pm0.06$ & 0.10\\ 
			PFL & $\log{(|B_{\mathrm{LOS}}|/\mu)}$  &$C$& $1.086\pm0.002$ & $0.010\pm0.001$ & $4.60\pm0.06$ & 0.14\\ 
			PFL & $\log{(|B_{\mathrm{LOS}}|/\mu)}$  &bisector& $1.064\pm0.002$ & $0.015\pm0.001$ & $3.93\pm0.06$ & 0.01\\ 
			LFL & $\log{(|B_{\mathrm{LOS}}|/\mu)}$  &$|B_{\mathrm{LOS}}|/\mu$& $0.653\pm0.003$ & $0.311\pm0.002$ & 1.00 & 2.24\\ 
			LFL & $\log{(|B_{\mathrm{LOS}}|/\mu)}$  &$C$& $0.578\pm0.003$ & $0.385\pm0.002$ & 1.00 & 4.89\\ 
			LFL & $\log{(|B_{\mathrm{LOS}}|/\mu)}$  &bisector& $0.622\pm0.003$ & $0.347\pm0.002$ & 1.00 & 3.31\\ 
			\hline\hline	
	\end{tabular}}
\tablefoot{Columns are: fit function, $x$ used in Eq. \ref{eq:logfit1}, the quantity over which the binning of the data was performed, best fit parameters ($a_1$, $a_2$, and $a_3$) with their 1$\sigma$ uncertainties, and the $\chi^2$ of the fits.}
\end{table*}

We first considered the data without the corrections for stray-light and without performing any segmentation other than excluding the sunspot regions. 
Figure \ref{fig:fitexample} shows the relation between the Ca~II~K brightness and $|B_{\mathrm{LOS}}|/\mu$ for all pairs of the degraded magnetograms and corresponding Ca~II~K images considered in our study.  
Each colour-coded pixel represents the logarithm of the number density within bins of 1 G and 0.01 in contrast.
The sources of the scatter seen in Fig. \ref{fig:fitexample} are discussed in more detail in Appendix \ref{sec:spatialagreement}. Briefly, one reason for the scatter of values is the projection effect.  SDO/HMI and Rome/PSPT observations sample different formation heights, which introduces changes in the distribution and shape of flux elements over space. Due to the expansion of the flux tubes with height, sizes of magnetic features at the two heights are different which leads to a size mismatch between the same feature seen in a magnetogram and in the corresponding Ca II K data and therefore also contributes to the scatter. Another source of scatter is the diverse spatial and spectral resolution of the compared data.
In Appendix \ref{sec:spatialagreement}  we discuss the spatial correspondence between the features in the magnetograms and the Ca II K observations and show close-ups of a quiet and an active region to demonstrate the smearing of the features in the Ca II K observations compared to the magnetograms.

The Spearman correlation coefficient between $|B_{\mathrm{LOS}}|/\mu$ and Ca~II~K contrast supports a monotonous relationship. The coefficient obtained for individual images is on average $\rho=0.60$, while it is $\rho=0.98$ for all pixels from all data. The significance level is zero with double-precision accuracy, implying a highly significant correlation. 

Figure \ref{fig:fitexample} shows that the Ca~II~K contrast increases with increasing magnetic field strength, but tends to saturate at high $|B_{\mathrm{LOS}}|/\mu$ \citep[see e.g., ][]{saar_empirical_1987,schrijver_relations_1989}.
The yellow curve in Fig. \ref{fig:fitexample} is a running mean over $|B_{\mathrm{LOS}}|/\mu$ values. Fitting the points of this binned curve has been the most common approach in the literature when studying the relation between Ca II K contrast and magnetic field strength \citep[e.g.][]{rast_scales_2003,ortiz_how_2005,rezaei_relation_2007,loukitcheva_relationship_2009,pevtsov_reconstructing_2016,kahil_brightness_2017,kahil_intensity_2019}. It suggests that the relation saturates at around 400 G. However, binning the data over the Ca~II~K contrast values suggests a somewhat different relation. 
We found that the choice of the quantity over which the binning is performed affects the exact form of the relation between the magnetic field strength and the Ca~II~K intensity. 
Attenuation bias due to errors in the independent variable in each case can cause these relations to skew compared to the true relationship.
This result, not reported in the literature yet, also needs to be considered when comparing outcomes from different studies.
We note that the histograms shown in Fig. \ref{fig:fitexample} illustrate that the distribution of contrast values is, to a good approximation, symmetric around the mean value for 150 G <$|B_{\mathrm{LOS}}|/\mu$< 450 G. The distribution for high and low $|B_{\mathrm{LOS}}|/\mu$ is skewed with a tail for high and low contrasts, respectively.

\begin{figure*}[t!]
	\centering
	\includegraphics[width=1.0\linewidth]{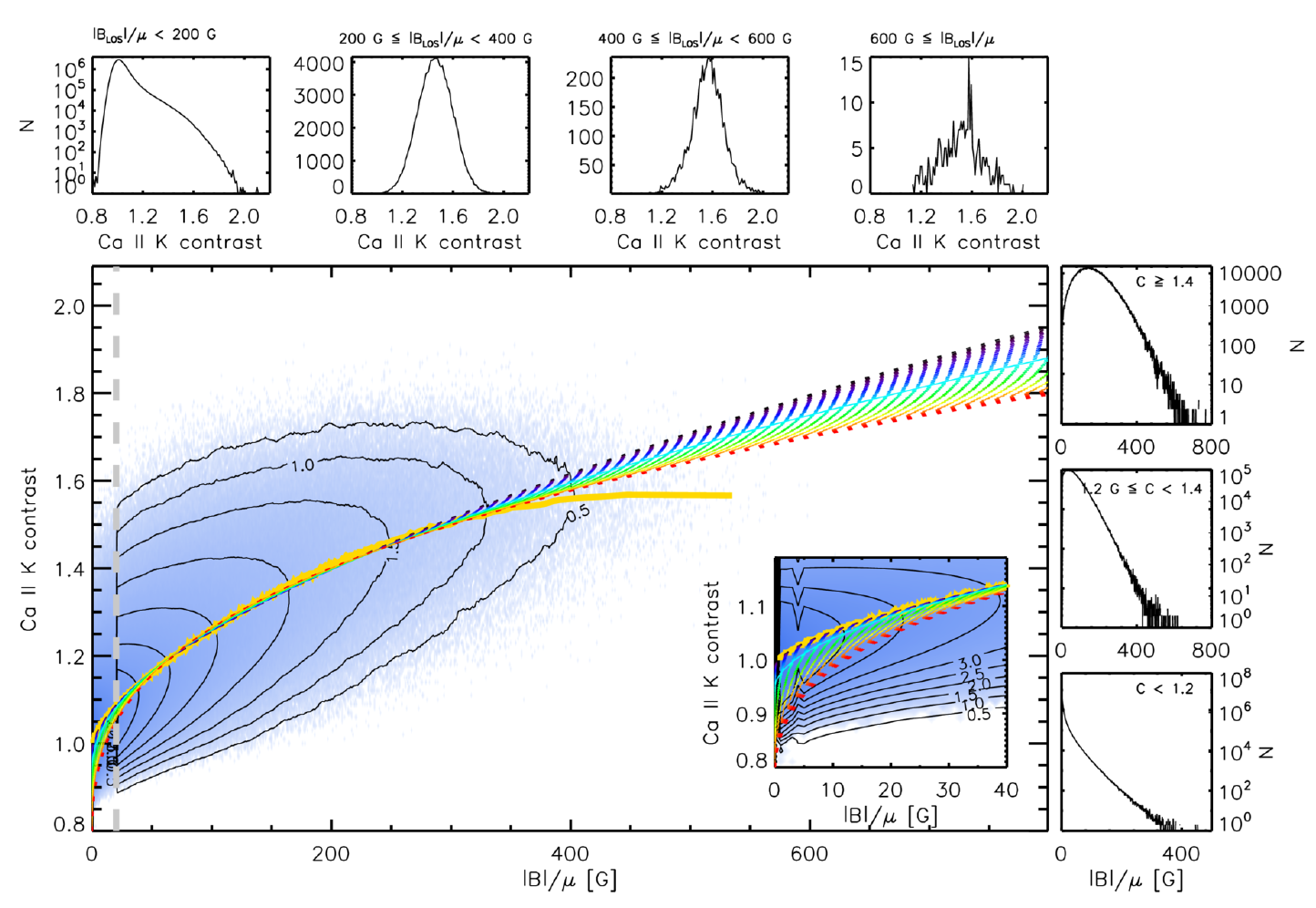}
	\caption{Same as Fig. \ref{fig:fitexample} but showing only the results of a power law function (PF) on the binned curve over $|B_{\mathrm{LOS}}|/\mu$ (yellow curve) by varying the $|B_{\mathrm{LOS}}|/\mu$ threshold, i.e., the magnetogram noise cut-off (dotted coloured curves). The threshold is 1 G for the black curve and gets to 50 G for the red curve. The curve corresponding to the 20 G threshold adopted in this study is shown with the light blue solid curve.
		The thick yellow curve shows 5000-point running mean over $|B_{\mathrm{LOS}}|/\mu$.  The vertical grey dashed line denotes the 20 G threshold in $|B_{\mathrm{LOS}}|/\mu$.}
	\label{fig:fitexample_lowlimit}
\end{figure*}

To find the best relation describing the data, we considered three different functions: a) a power law with an offset (PF) as commonly used in the literature \citep[e.g.][]{schrijver_relations_1989,harvey_magnetic_1999,ortiz_how_2005,rezaei_relation_2007,loukitcheva_relationship_2009}; b) a  logarithm (LFL) as proposed by \citet{kahil_brightness_2017,kahil_intensity_2019}; and c) a power law function of the logarithm of $|B_{\mathrm{LOS}}|/\mu$ (PFL). These three functions can be described by the following equation:

\begin{equation}
C=a_1+a_2x^{a_3},
\label{eq:logfit1}
\end{equation}
where $x=|B_{\mathrm{LOS}}|/\mu$ for PF, and $x=\log{(|B_{\mathrm{LOS}}|/\mu)}$ for PFL and LFL (with $a_3$ being 1 for LFL).
We perform these fits on the curve that resulted by averaging contrast values over $|B_{\mathrm{LOS}}|/\mu$ values (yellow curve in Fig. \ref{fig:fitexample}), based on all selected pixel pairs from all images where $|B_{\mathrm{LOS}}|/\mu \ge 20$ G. However, for comparison we also performed the fits on the curve after binning over contrast values and on the bisector of the two running means (these sets of fits will be referred to as PF$^*$, PFL$^*$, and LFL$^*$). 

The fits with the three tested functions for the $|B_{\mathrm{LOS}}|/\mu$ binning (yellow solid line in Fig. \ref{fig:fitexample}) are shown in Fig. \ref{fig:fitexample}, with red dashed line (PF), blue dotted line (PFL), and green dashed line (LFL). Table \ref{tab:functionparameters} lists the derived parameters. 

Both PF and PFL give low values for $\chi^2$, being $\simeq 0.16$ and $\simeq 0.1$, respectively. The fitted curve for both PF and PFL does not follow the binned curve for high $|B_{\mathrm{LOS}}|/\mu$, lying above it. 
PF and PFL closely follow each other up to about 400 G, but slightly diverge at higher magnetic field strengths, with PFL following the binned curve more closely. They also differ for $|B_{\mathrm{LOS}}|/\mu<20$ G (which were not included in the fit, so that the curves are extrapolated there), with PFL giving higher contrasts. However, the differences between the two curves are minute.
We found that the exponents for PF and PFL increase when the fit is performed on the curves  binned over contrast values or on the bisector (see Table \ref{tab:functionparameters}), while the $\chi^2$ is reduced, being 0.03 and 0.01, respectively. 
LFL fails to reproduce the binned curve over $|B_{\mathrm{LOS}}|/\mu$, but follows slightly better the trend of the curve for $|B_{\mathrm{LOS}}|/\mu>400$ G than PF and PFL.
However, the fit of LFL gives high values for $\chi^2$ (2.24), showing that LFL does not describe the data well.

The analysis described in the following was performed by applying all functions and binning curves described above to the available data. However, due to the similarity of the results obtained from the PF and PFL fits and the lower accuracy of the LFL fit compared to both PF and PFL fits, for the sake of clarity we will present only the results for PFL$^*$ and PF. Our analysis suggests that the PFL$^*$ fit is more accurate and stable (see Sec. \ref{sec:reconstructing_magnetograms}) than the other considered functions, 
while those derived with the PF fit allow for  comparison with previous results in the literature.
We note, however, that due to the scatter in Fig. \ref{fig:fitexample} we cannot rule out the aptness of PF to describe the relation between the magnetic field strength and the Ca II K brightness either.
The results derived with PF$^*$ and LFL$^*$ fits can be found in \cite{chatzistergos_analysis_2017}.

\subsection{Effects of the $|B_{\mathrm{LOS}}|/\mu$ threshold on the derived exponents}
To better understand the sources of differences with other results, we have studied how our findings depend on the $|B_{\mathrm{LOS}}|/\mu$ threshold applied. Figure \ref{fig:exponents_wholediscfeatures_lowthreshtest} shows the parameters derived by applying PF and PFL to the data shown in Fig. \ref{fig:fitexample} and varying the threshold in $|B_{\mathrm{LOS}}|/\mu$ between 1 G and 50 G, i.e. 0.15 to 7 times the noise level. We show only the exponents, though the other parameters of the tested functions are affected as well. We show the results of performing the fit to all three binned curves as shown in Fig. \ref{fig:fitexample}. For the binning over $|B_{\mathrm{LOS}}|/\mu$, the exponent for PF constantly decreases, while for PFL it reaches a plateau for magnetic field strengths in the range of $\sim$5--20 G and then slightly decreases. 
When the fit is performed on the bisector, the exponents for PF$^*$ and PFL$^*$ reach a plateau after a threshold of $\sim$8 G and $\sim$18 G, respectively, and after that they tend to slightly decrease. The exponent we derived for PFL$^*$ (Table \ref{tab:functionparameters}) lies within the 1$\sigma$ interval of all derived exponents with thresholds greater than 18 G.
For the binning over contrast values, the exponents of PF and PFL show an almost constant increase for $|B_{\mathrm{LOS}}|/\mu>10$ G. 

\begin{figure}[t!]
	\centering
	\includegraphics[width=1\linewidth]{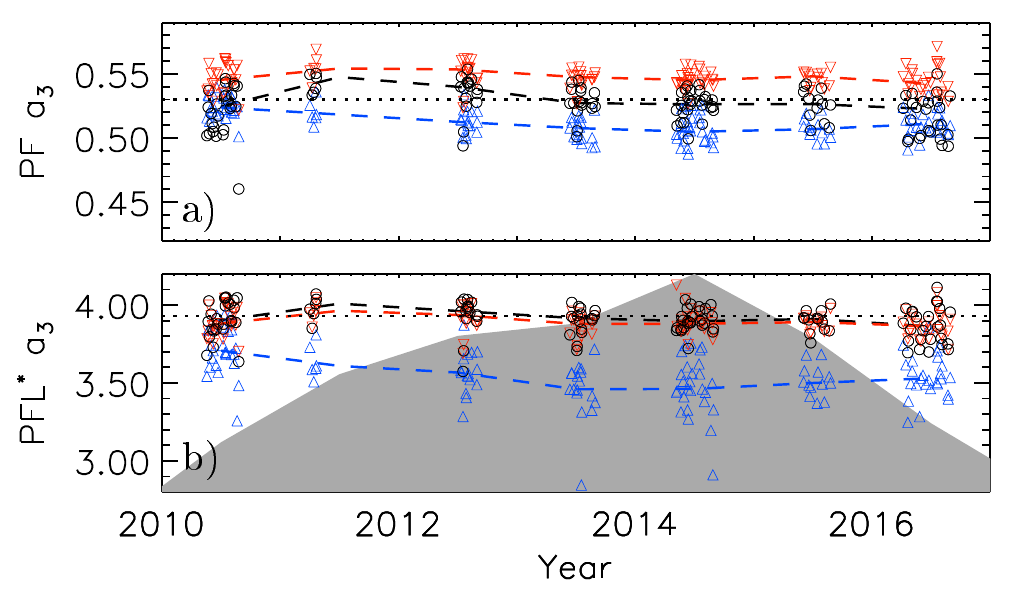}
	\caption{Parameters of the fits (Eq. \ref{eq:logfit1}) as a function of time, derived for all bright features (black), as well as for the network (blue), and plage (red) separately, for PF (a)) and PFL$^*$ (b)) fits. The dashed lines connect the median values obtained from all analysed images within a given year, while the dotted lines mark the values of the parameters of the best fit derived in Sect. \ref{sec:scatterpots}. The shaded grey surface in panel b) shows the plage areas determined with Method 1 from the Rome/PSPT images. The areas were scaled to have a maximum value of 4.2 and minimum value 2.8 in order to match the range of values shown in panel b). They indicate the level of solar activity at the considered times. }
	\label{fig:exponents_wholediscfeatures}
\end{figure}

\begin{figure}[t!]
	\centering
	\includegraphics[width=1\linewidth]{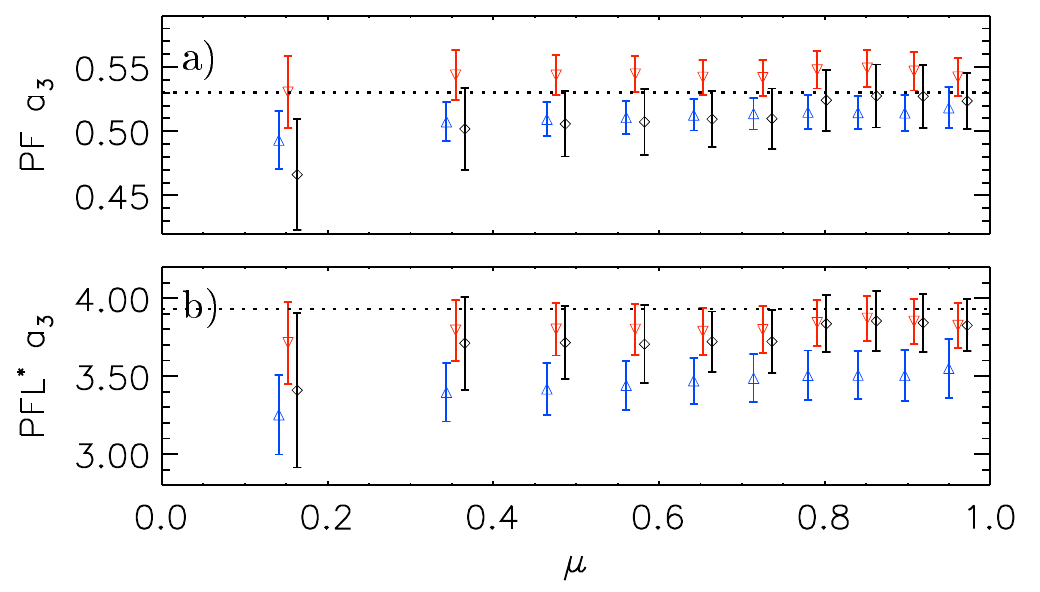}
	\caption{Parameters of the fits (Eq. \ref{eq:logfit1}) as a function of $\mu$, derived for 10 annuli of equal area for all bright features (black), as well as for the network (blue), and plage (red) separately, for PF (a)) and PFL$^*$ (b)).
		The values shown are the means over the entire sample of data, while the error bars denote the $1\sigma$ interval. Results for the network are shown in the middle of the $\mu$ interval they represent, while the others are slightly shifted in $\mu$ to improve the clarity of the plot. The dotted lines mark the values of the best fit parameters derived in Sect. \ref{sec:scatterpots}}
	\label{fig:exponents_annulidiscfeatures}
\end{figure}

\begin{figure}[t!]
	\centering
	\includegraphics[width=1\linewidth]{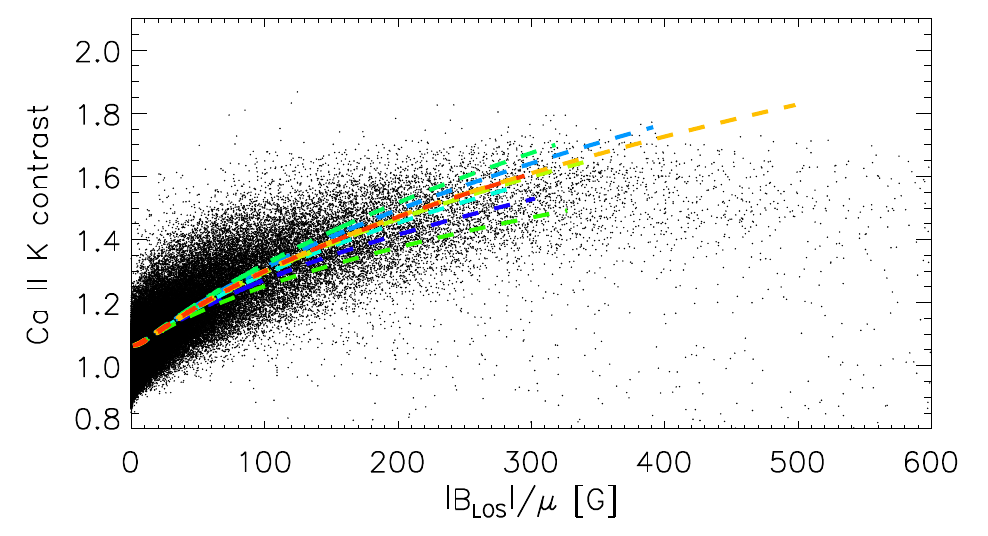}
	\caption{Ca~II~K contrast plotted against $|B_{\mathrm{LOS}}|/\mu$ for all activity clusters identified with Method 2 in observations shown in Fig. \ref{fig:psptcaexamples} (black dots). The coloured curves are the result of the PFL$^*$ fit to the individual clusters, shown with the same colours as in Fig. \ref{fig:segmentationmasks}c).} 
	\label{fig:20110401090600bversuskplage}
\end{figure}

\begin{figure}[t!]
	\centering
	\includegraphics[width=1\linewidth]{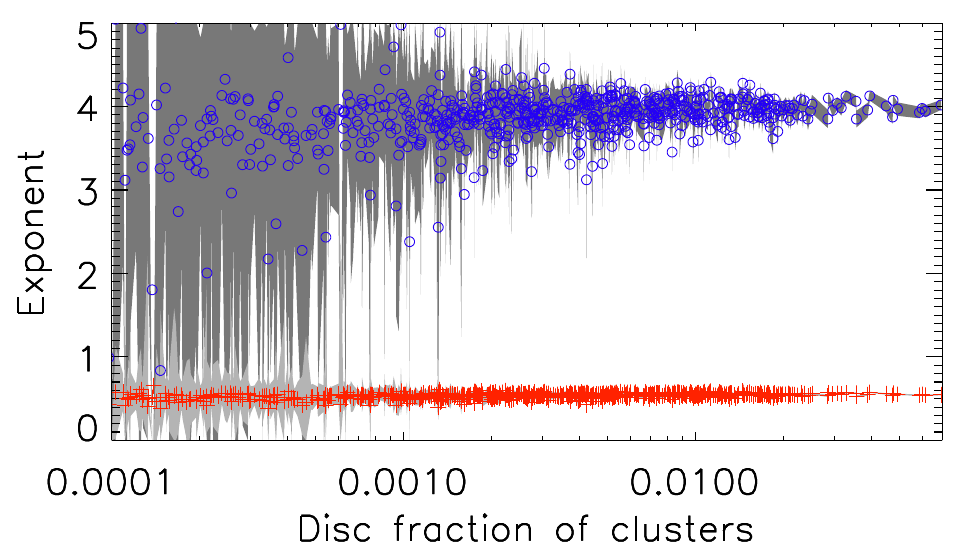}
	\caption{Exponents of the PF (red) and PFL$^*$ (blue) fits as a function of size for individual activity clusters. The light grey and dark grey shaded areas denote the 1$\sigma$ error in the fit parameters for PF and PFL$^*$, respectively.}
	\label{fig:ARexponentwithsize}
\end{figure}

\begin{figure}[t!]
	\centering
	\includegraphics[width=1\linewidth]{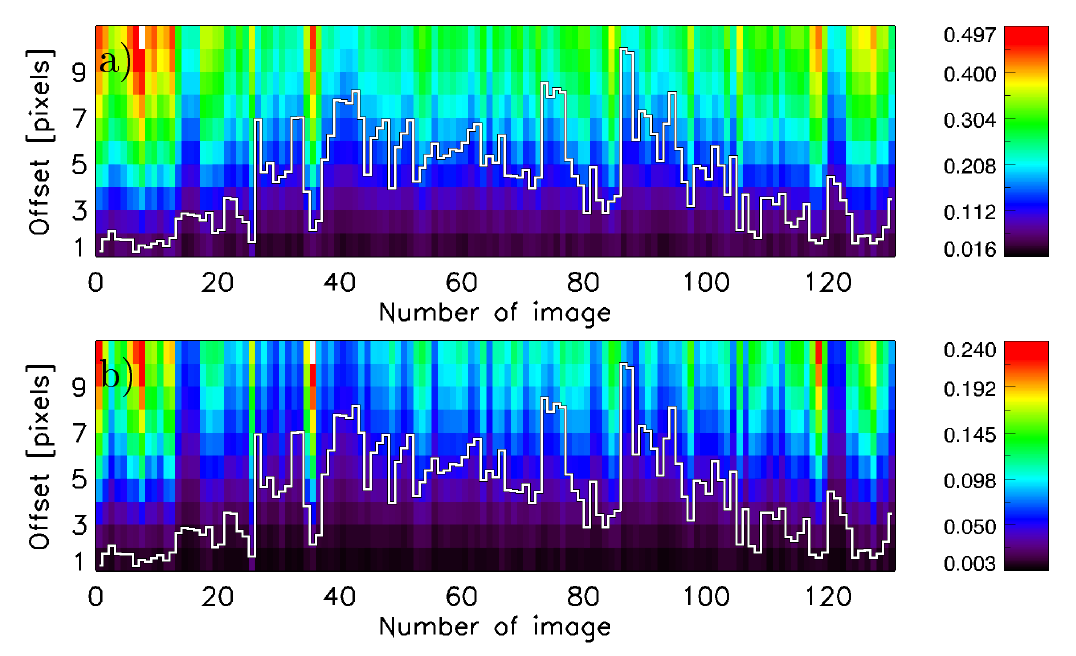}
	\caption{Colour-coded relative errors in the exponent derived with PF (a)) and PFL$^*$ (b)) due to misalignment of the analysed images. Each Ca~II~K  image was randomly shifted in both $x$ and $y$ directions by up to 10 pixels. The boxes give the average error after 1000 realisations. The $y$ axis gives the value of the maximum possible offset in any direction. The $x$ axis indicates the analysed images ordered by date covering the period 2010--2016. Colour bars show the relative errors in the computed exponents. Over-plotted with white are the plage areas derived from the Ca II K images. The areas were scaled in the range [0,9] and are shown to indicate the activity level.}
	\label{fig:misalignmenterrors}
\end{figure}

The threshold seems to play a more important role if the fit is performed by binning over contrast values or $|B_{\mathrm{LOS}}|/\mu$ compared to the results of the fit on the bisector. Overall, the curves derived with PFL$^*$ are more stable against the choice of the $|B_{\mathrm{LOS}}|/\mu$ threshold. 
In Figure \ref{fig:fitexample_lowlimit} we show the results of fitting PF to the binned curve over $|B_{\mathrm{LOS}}|/\mu$ by varying the threshold between 1 G and 50 G. All of the derived curves agree very well for the interval 50--350 G, but they diverge for higher and lower values.
This is expected since the low $|B_{\mathrm{LOS}}|/\mu$ regions dominate the relation and by increasing the threshold it shifts the weight for the fit to higher $|B_{\mathrm{LOS}}|/\mu$.

\subsection{Exponents over time and different $\mu$ positions}
\label{sec:caexponentstimemu}

We also studied if the exponents of the fits change with the activity level. To understand the change with time we performed the fits on every image separately, first for all pixels with $\mu>0.14$, and then for the plage and network regions separately. The differentiation between the various types of  features was done  with Method 1 applied to Rome/PSPT images, but keeping only the regions that also have $|B_{\mathrm{LOS}}|/\mu>20$ G in the magnetograms.
To study the variation of the exponent we fixed $a_1$ and $a_2$ for the PF and PFL fits to the values derived in Sect. \ref{sec:scatterpots} (listed in Table \ref{tab:functionparameters}). 

Figure \ref{fig:exponents_wholediscfeatures} shows the coefficients of the fits to the curve binned over $|B_{\mathrm{LOS}}|/\mu$ as a function of time. 
The resulting exponents for PF and PFL depend on the type of feature and are slightly higher for plage than for the network. 
The uncertainty in the derived parameters (not shown in Fig. \ref{fig:exponents_wholediscfeatures} due to their low values) is less than 0.001 for $a_3$ in PF and 0.014 for $a_3$ in PFL$^*$. 
Performing the fit to all pixels on the disc with $\mu>0.14$ and each image separately, we found an average exponent of $0.52\pm 0.02$ and $3.9\pm 0.1$ for PF and PFL$^*$, respectively. 
The errors are the $1\sigma$ intervals among all the daily calculated values. These values agree within the $1\sigma$ uncertainty level with those we derived in Sect. \ref{sec:scatterpots} for all three functions. As seen in Fig. \ref{fig:exponents_wholediscfeatures} the scatter of the resulting exponents is such that within the limits of the current analysis, we found no evidence that the relationship between $|B_{\mathrm{LOS}}|/\mu$ and Ca~II~K intensity varies over the solar cycle. We noticed exactly the same behaviour for the plage component for PF and PFL$^*$. 
We found some changes in the network component that result in higher exponents for the low activity period in 2010 for PF and PFL$^*$, but the derived exponents are still constant in time within the uncertainties.

\begin{figure*}[t!]
	\centering
	\includegraphics[width=1\linewidth]{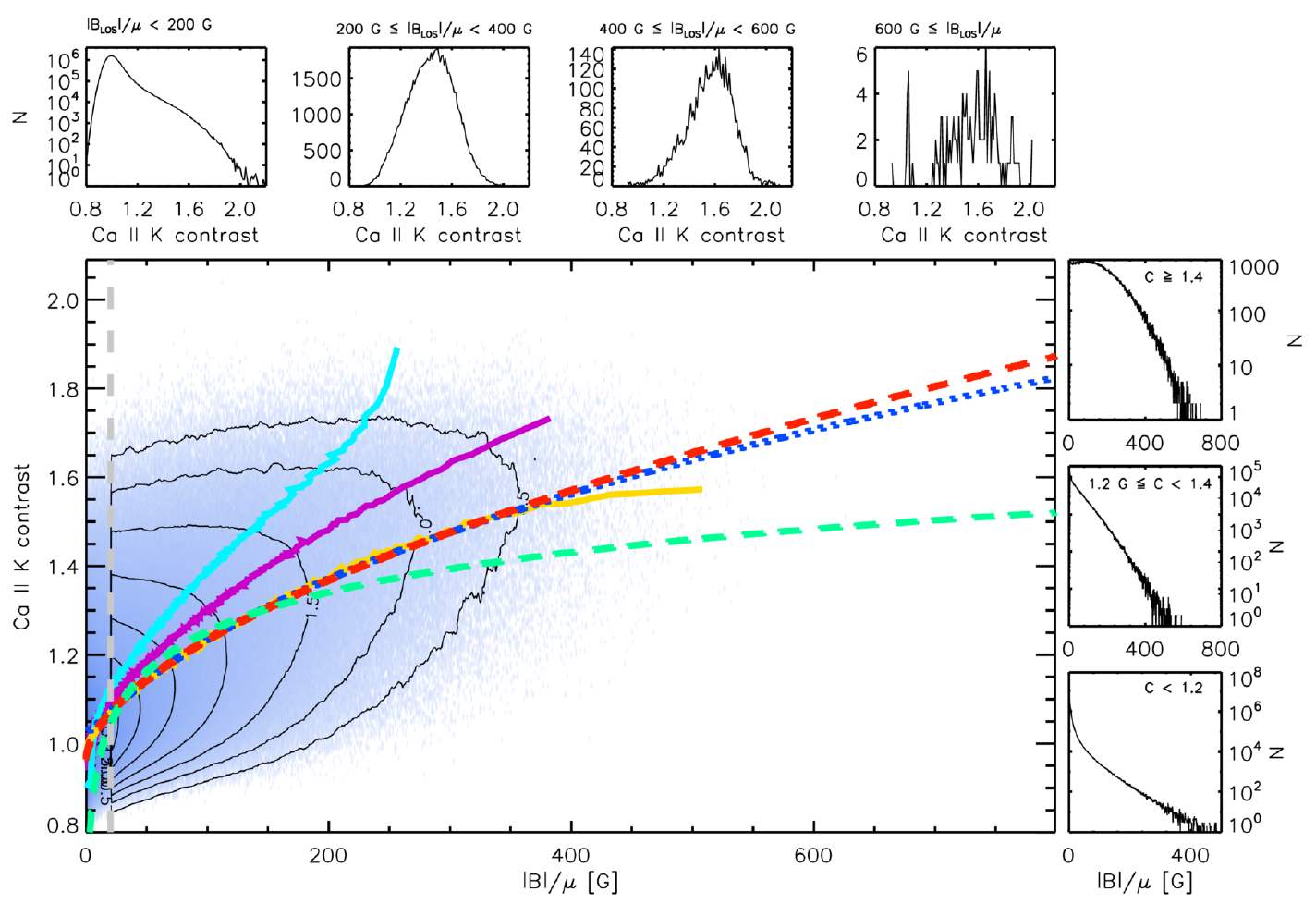}
	\caption{Same as Fig. \ref{fig:fitexample} but for stray-light corrected data. }
	\label{fig:fitexample_slc}
\end{figure*}

\begin{figure*}
	\centering
	\includegraphics[width=1\linewidth]{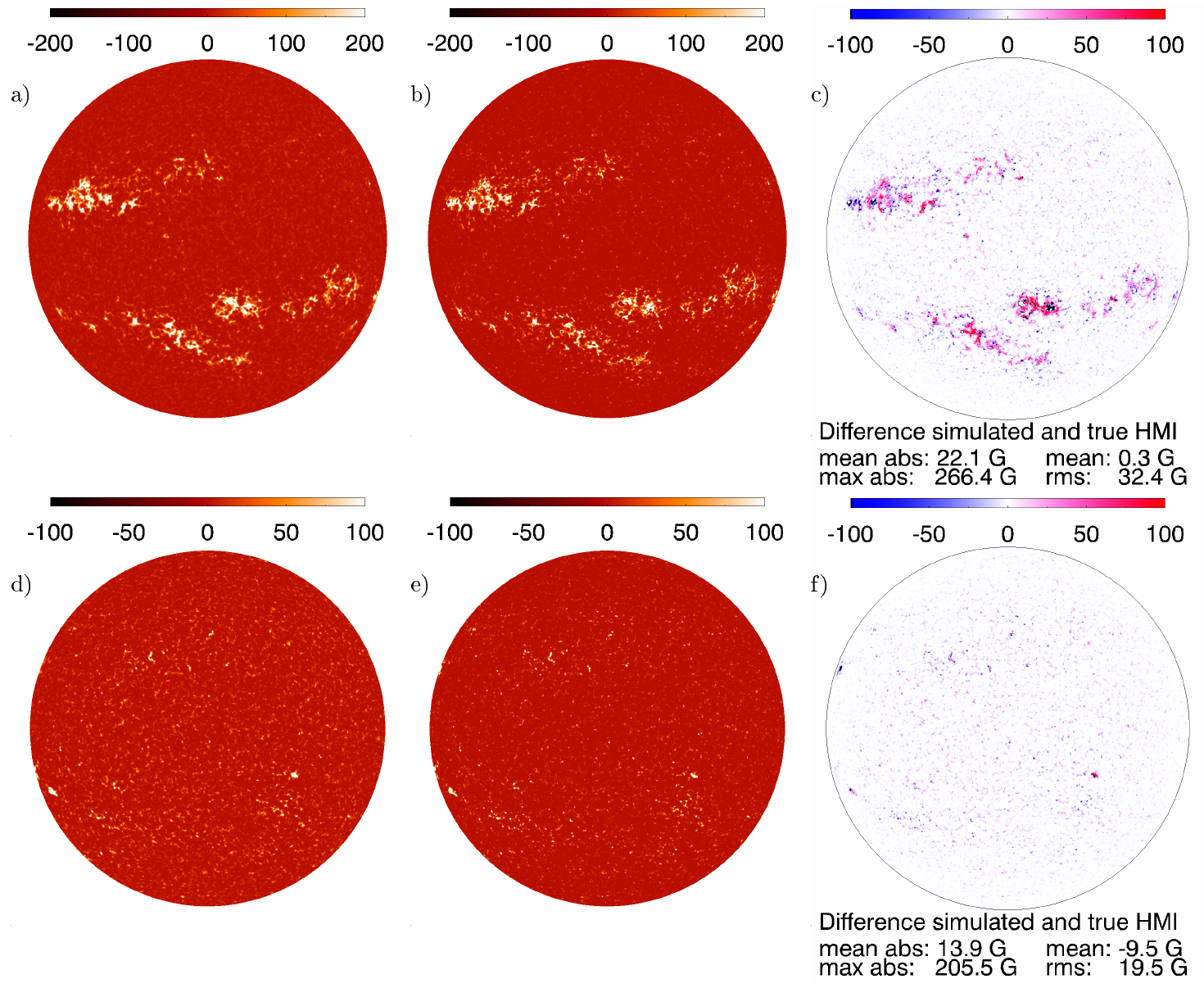}
	\caption{Unsigned magnetograms reconstructed from the Ca~II~K images taken on 01/04/2011 (top) and 07/06/2010 (bottom) using the average parameters for PFL$^*$ (left), SDO/HMI unsigned magnetograms  (middle) co-temporal to the Ca II K images, and difference between the reconstructed unsigned magnetogram (simulated) from Ca~II~K data and the original (true) SDO/HMI unsigned magnetogram (right). The RMS, mean, mean absolute, and maximum absolute differences are listed under each panel. The colour bars show the ranges of $|B_{\mathrm{LOS}}|$ in G. All images are saturated at 100 G to improve the visibility of the regions with low magnetic field strength}
	\label{fig:20000821nsb}
\end{figure*}

We have also studied how the exponents of the fits change for different positions of the disc.
Figure \ref{fig:exponents_annulidiscfeatures} shows the  coefficients of the various features 
as a function of their position on the solar disc in terms of $\mu$. The segmentation was done with Method 1, identifying plage and network regions. We considered  10 concentric annuli of equal area covering the solar disc up to $\mu=0.14$. 
The mean values of the exponents computed over the various annuli slightly decrease towards the limb, but their standard deviation increases so that the exponents do not show any significant variation with the position on the disc (within the 1$\sigma$ uncertainty). In particular, the relative difference between the average value of the exponents within the innermost and outermost annuli for PF (PFL$^*$) is 4\% (10\%). The same behaviour is seen also when network and plage regions are considered separately.

\subsection{Exponents for individual activity clusters}

We have also tested how different the exponents of the fits are when applied to the data from individual activity clusters. The images were segmented with Method 2. We performed the fit with PFL$^*$ to each individual cluster, while we also considered the QS (including the network) separately. 
Figure \ref{fig:20110401090600bversuskplage} shows a scatter plot for the images shown in Fig. \ref{fig:psptcaexamples}, but now including only the pixels corresponding to activity clusters and QS. The binned curves and the result of the fits from different activity clusters are in agreement with each other, with the exception of one cluster. However, this cluster is very small in size and the statistics are worse than for the other clusters.
The relation derived from QS regions shows a smaller slope than the one obtained for active regions. However, this is probably due to a much lower number of QS and network pixels with strong resolved magnetic fields in the analysed SDO/HMI degraded data. 

Results for different clusters agree well with each other within the accuracy of the fit. 
Averaging all exponents derived for clusters (QS and network) from all images gave on average the values of $0.54\pm0.03$ ($0.50\pm0.01$), and $3.9\pm0.2$ ($3.7\pm0.1$), for PF and PFL$^*$, respectively.
The exponents derived here are in agreement with those presented in the previous subsections. 
We find no dependence on $\mu$ for the derived exponents with this segmentation method either.

Figure \ref{fig:ARexponentwithsize} shows the exponents derived with PF (red) and PFL$^*$ (blue) as a function of the area of the clusters expressed in fractions of the disc. We found no dependence of the exponent on the cluster size, however the uncertainty of the derived parameters was obviously higher for smaller features, because of poorer statistics. 
Also, effects of potential misalignment between SDO/HMI and Rome/PSPT data become more significant in this case. Find more information in Sect. \ref{sec:misalignment}.

\subsection{Effects of potential misalignment}
\label{sec:misalignment}
To test the sensitivity of our results to potential misalignment of the images, we repeated our analysis by using Rome/PSPT images shifted by a random number of pixels, in both the $x$ and $y$ direction and compared the results with those from the original Rome/PSPT images. 
The test was done 10 times, whereby the possible maximum offset varied between 1 and 10 pixels (2'' and 20'') in any direction. Each time, we performed 1000 computations with a random offset for each image lying in the range between 0 and the maximum allowed value. The choice of a maximum offset of 10 pixels is extreme considering that the alignment with the data employed in this study is considerably more precise. However, it is useful to test such a high value in order to have an estimate of the errors when applying the relationships on lower resolution data or with greater temporal difference than the images analysed here, such as the magnetograms from Kitt Peak and Ca II K spectroheliograms from Kodaikanal observatory, which are taken on average more than 12 hours apart. Figure \ref{fig:misalignmenterrors} shows the relative difference of the derived exponents on each offset image to the original ones. Shown are the average values over the 1000 realisations on each image (abscissa) for each maximum possible offset (ordinate). We show the errors for the exponents derived for PF and PFL$^*$ in different panels.
We notice that the errors are significant for PF, but they are considerably lower for PFL$^*$. The errors in the derived parameters with PF reach 50\%, while they are less than 8\% for 1 pixel offsets. With offsets up to 10 pixels (20''), the errors for the derived parameters with PFL$^*$ remain below 24\%, while they are less than 13\% and 2\% for 5 and 1 pixel offsets (10'' and 2''), respectively. We noticed that errors due to the offsets are higher during low activity periods with all tested functions. 
This may be due to the smaller size of individual magnetic features when activity is low, so that an offset quickly leads to a substantial mismatch between the magnetic features in SDO/HMI and the brightness features in Rome/PSPT images.

\subsection{Effects of stray-light}
We studied the effect of the stray-light on our results. For this, we repeated the same analysis on images corrected for stray light (as described in Sect. \ref{sec:straylight}). Since the Ca II K stray-light corrected images have higher contrast values, the segmentation parameters for different features had to be adapted (increased by 0.02 in contrast and 10 G). Otherwise the methods that we applied were exactly the same. Figure \ref{fig:fitexample_slc} is similar to Fig. \ref{fig:fitexample}, but now for the stray-light corrected data. The scatter in Ca~II~K contrast is higher compared to the one from the Ca II K images affected by stray-light.
However, our results remain unchanged with almost constant exponents over the disc and time, and values of the exponents close to the ones reported above. The best fit parameters from images corrected for stray-light are $0.51\pm 0.02$ and $3.89\pm 0.08$ for PF and PFL$^*$, respectively. 
Besides, there are no significant differences to the results obtained from analysis of the data corrected with the method of \cite{criscuoli_photometric_2017}, thus supporting the assumptions of the correction method by \cite{yeo_point_2014}.  
Our previous conclusions of weak CLV of the exponents and time independence are also valid with these data. The values of the exponents are slightly lower than in the rest of our analysis, with best fit parameters $0.50\pm 0.03$ and $3.88\pm 0.05$ for PF and PFL$^*$, respectively. These results are still within the 1$\sigma$ interval of our main results based on stray-light uncorrected data.

\subsection{Comparison to results from the literature}
\label{sec:somparisontoothers}
The exponent derived for PF ($0.53\pm0.01$) is lower than those obtained by \cite{schrijver_relations_1989,harvey_magnetic_1999,ortiz_how_2005}, who favoured an exponent of 0.6, 0.69, 0.65, respectively, for all bright features considered. However, it is higher than those derived by \cite{rezaei_relation_2007,loukitcheva_relationship_2009} and \cite{vogler_solar_2005}, ranging from 0.31 to 0.51. The difference between our results and those by \cite{loukitcheva_relationship_2009} (exponent of 0.31) can potentially be explained by the different threshold in the magnetic field strength used by the two studies.
\cite{rezaei_relation_2007} found the exponent to increase to 0.51 when the threshold was 20 G, which is consistent with our results.
The same stands for \cite{loukitcheva_relationship_2009} who used a threshold of 1.5 G, and showed that the exponent increases to roughly 0.53 if a threshold of 20 G is used. 
It is worth noting, however, that \cite{loukitcheva_relationship_2009} analysed lower resolution  magnetograms from the Solar and Heliospheric Observatory Michelson Doppler Imager magnetograms \citep[SOHO/MDI][]{scherrer_solar_1995}, while we analysed SDO/HMI magnetograms and hence the magnetic field strengths reported by the various instruments is not necessarily directly comparable \citep[c.f.][]{yeo_reconstruction_2014}.
We note that the exponents derived by \cite{schrijver_relations_1989,harvey_magnetic_1999,ortiz_how_2005} are consistent with the one we derive here by performing the fit on the bisector (PF$^*$).

Our results for LFL differ from those presented by \citet{kahil_brightness_2017} and \citet{kahil_intensity_2019}. For the LFL \citet{kahil_brightness_2017,kahil_intensity_2019} reported the best fit parameters $a_1=0.29\pm0.003$ and $a_2=0.51\pm0.004$ for $|B_{\mathrm{LOS}}|>50$ G, and  $a_1=0.456\pm0.003$ and $a_2=0.512\pm0.001$ for $|B_{\mathrm{LOS}}|>100$ G for a QS and an AR in Ca II H data, respectively. 
The differences can be due to different atmospheric heights sampled in the analysed data, as well as due to the lower spatial resolution of the observations used here compared to those used by \cite{kahil_brightness_2017,kahil_intensity_2019}. 
Thus, LFL might be not an appropriate function for analysis of full-disc data like those used in our study.

\cite{harvey_magnetic_1999} analysed data from three observatories, specifically the Big Bear, Sacramento Peak, and Kitt Peak observatories, and segmented the features into 4 sub-categories.
In addition to the categories we use, they also have a feature class they termed enhanced network. Our results are close to those of \cite{harvey_magnetic_1999} for the Big Bear data (0.52 and 0.58 for plage and network, respectively), but are slightly higher than those from Sacramento Peak  (0.47--0.48 and 0.47--0.56 for plage and network, respectively) and lower than those of Kitt Peat (0.62 and 0.64 for plage and network, respectively) measurements. This can potentially be explained by the different bandwidth of the observations made at the different observatories. Indeed, the Big Bear data have a bandwidth of 3~\AA~being the closest to the one of the Rome/PSPT (2.5~\AA). The bandwidth used for the Sacramento Peak data is narrower (0.5~\AA), while for the Kitt Peak data it is broader (10~\AA). Another difference is that \cite{harvey_magnetic_1999} found lower (or equal) exponents for the active regions than for the network, while in our study we found the opposite. 
Note that the exponent obtained for the enhanced network component by \cite{harvey_magnetic_1999} is higher than the one we derived here for network and plage. Our finding of small to no dependence of the exponents to $\mu$ is in agreement to the results of \cite{harvey_magnetic_1999}.

\cite{pevtsov_reconstructing_2016} analysed the pairs of Kitt Peak magnetograms and uncalibrated Mt Wilson Ca~II~K spectroheliograms after converting them to carrington maps, 
as well as SOLIS/VSM observations in Ca II 854.2 nm and magnetograms. They concluded that Ca II brightness is an unreliable proxy for the magnetic field strength, 
because of the large scatter between Ca~II~K brightness and the magnetic flux and that they saw a reversal of the relationship at high magnetic fluxes.
It should be noted, however, that the data they analysed were of significantly lower quality than the ones we used. This is manifested by the number of pixel pairs and years analysed: $\sim$62,000 over 12 years in \cite{pevtsov_reconstructing_2016} and $\sim$103,000,000 over 6 years of Ca II K data. 
The issue of the low spatial resolution of the Ca 854.2 nm line was mentioned by \cite{pevtsov_reconstructing_2016} based on the findings of \cite{leenaarts_comparison_2006}.
The reported reversal of the relation at high magnetic fluxes with the Ca II K data, as well as the lack of correlation for the Ca II infra-red data considered by \cite{pevtsov_reconstructing_2016}, is perfectly consistent with the inclusion of sunspots in their analysis. 
The large scatter is possibly due to the narrower nominal bandwidth of Mt Wilson data (0.35 \AA{}) compared to that of the Rome/PSPT (2.5 \AA{}). This means that Mt Wilson is sampling greater atmospheric heights than Rome/PSPT, where the flux tubes are more expanded and hence the spatial agreement between the Ca II K data and the magnetograms should be reduced. In addition, at greater height, the emitted radiation is more strongly affected by shock waves and local heating events, which reduce the agreement even more. 
Lack of photometric calibration of the historical images and potential not very accurate methods applied for processing can distort the relation too \citep[see][]{chatzistergos_analysis_2017,chatzistergos_analysis_2018,chatzistergos_analysis_2019}. 

\section{Reconstructing unsigned magnetograms from Ca~II~K images}
\label{sec:reconstructing_magnetograms}
In the previous section we showed that the exponents of the functions tested in our study remained time- and $\mu$ independent. This allows us to reconstruct unsigned magnetograms or pseudo-magnetograms from the full-disc Ca~II~K observations by using the parameters derived in Sect. \ref{sec:scatterpots}. 

For this, we apply the 3 tested relationships with  the best-fit parameters listed in Table \ref{tab:functionparameters} on the Ca II K observations. 
We used the parameters from all three different binning approaches. However, we noticed that using the bisector fit produced magnetograms with the lowest differences to the original ones. We also found that parameters derived from the $|B_{\mathrm{LOS}}|/\mu$ binning tend to result in magnetograms with overestimated bright regions compared to the original magnetograms. This is also found in magnetograms reconstructed with the fits to the bisector, however to a lesser degree. The magnetograms reconstructed with the parameters from the binning over contrast values tend to underestimate the magnetogram signal in large parts of the bright regions. Based on this, in the following we present the results for magnetograms obtained with the parameters for PFL$^*$ applied to the bisector. However, for comparison, we also show in the appendix \ref{sec:appendix_comparison} the results of PF$^*$, LFL$^*$, and PF.

The pixels with contrast $\leq1$ were set to 0 G. 
Figure \ref{fig:20000821nsb} shows examples of reconstructed magnetograms for an active and a quiet day by applying the best fit PFL$^*$ relationship on the Rome/PSPT Ca~II~K image (panels a), d)) and the corresponding SDO/HMI magnetogram (panels b), e)). The pixel by pixel absolute differences between the reconstructed and the original magnetograms are shown in Fig. \ref{fig:20000821nsb}  (panels c), f)). 
Prior to getting the differences, the original and reconstructed magnetograms were multiplied with $\mu$, so that the compared quantity is $|B_{\mathrm{LOS}}|$. 
In this reconstruction we only made use of the information on the Ca~II~K image to identify the regions on which we applied the relationships obtained in Sect. \ref{sec:scatterpots}. This means that sunspots were not identified accurately and their immediate surroundings were the regions with the highest errors, reaching differences of up to $\sim 1000$ G. These regions were masked out in Fig. \ref{fig:20000821nsb} and the errors reported in the plots do not include sunspots. 

Comparing the errors between the reconstructed and the original magnetograms 
we got RMS differences of $\simeq30$ G and $\simeq20$ G for the active and quiet day, respectively. The differences for the quiet day show that we slightly underestimated the weak fields.

\begin{figure}
	\centering
	\includegraphics[width=1\linewidth]{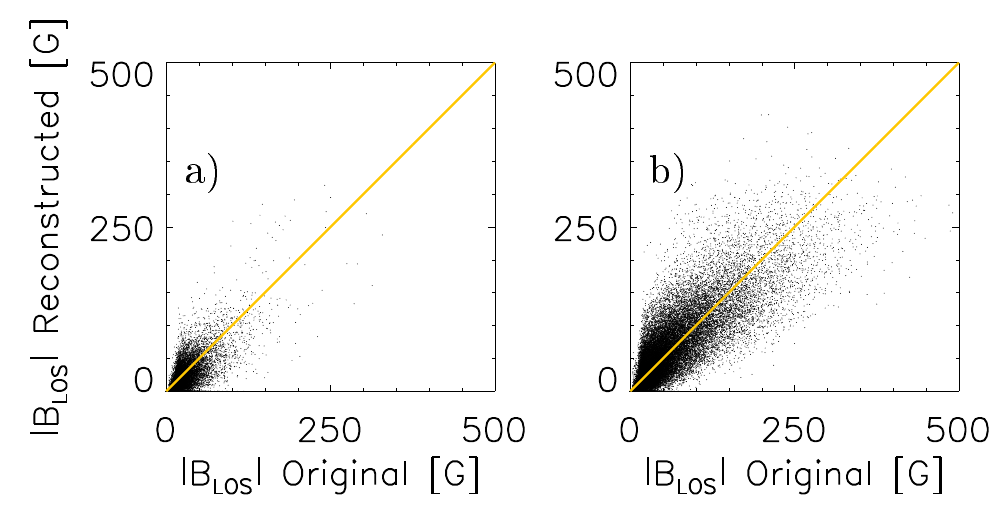}
	\caption{Scatter plots between original (degraded) magnetogram and the reconstructed from the Ca~II~K image taken on 07/06/2010 (a) and 01/04/2011 (b) using the average parameters for PFL$^*$. The yellow line has a slope of unity. The axes are shown in the range from the original magnetogram.}
	\label{fig:scatterplots_originalreconstructedmagnetograms}
\end{figure}

\begin{figure}
	\centering
	\includegraphics[width=1\linewidth]{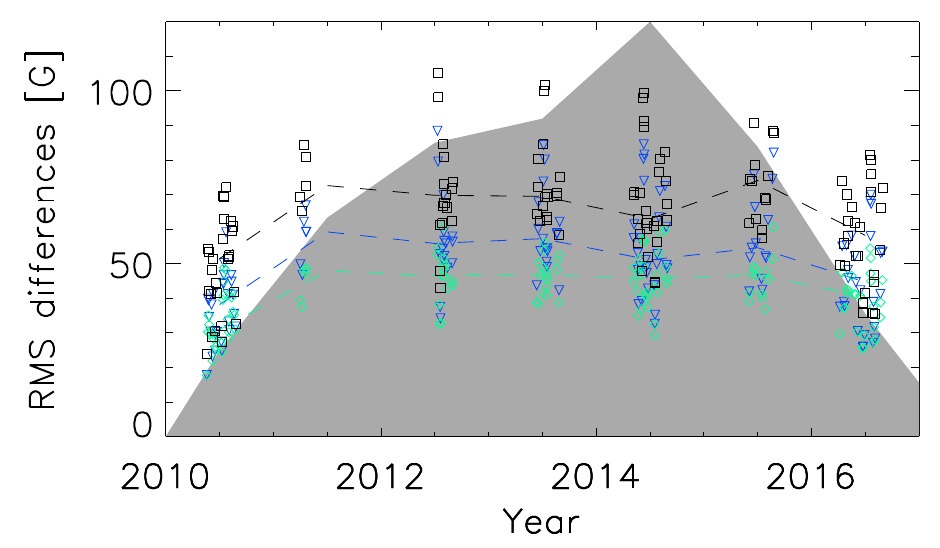}
	\caption{RMS pixel by pixel differences in G between the original unsigned magnetograms and the reconstructed ones using the parameters (listed in Table \ref{tab:functionparameters}) derived from the PFL$^*$ fits for the whole disc (blue downward triangles) and by masking out the sunspot regions (green rhombuses). Also shown is the standard deviation of the original unsigned magnetograms (black squares). The dashed lines connect annual median values. The shaded grey surface shows the plage areas determined with Method 1 from the Rome/PSPT images. The areas were scaled to have a maximum value of 120 and are included to indicate the level of solar activity. }
	\label{fig:differences25070}
\end{figure}

Figure \ref{fig:scatterplots_originalreconstructedmagnetograms} shows scatter plots between the reconstructed magnetograms and the original ones for the observations shown in Fig. \ref{fig:20000821nsb}. 
Figure \ref{fig:differences25070} shows the pixel by pixel RMS differences  between the original and the reconstructed unsigned magnetograms obtained using the derived best fit parameters of PFL$^*$, without masking the surroundings of the sunspots this time.
Figure \ref{fig:differences25070} reveals that the RMS differences remain less than 88 G for all 131 reconstructed unsigned magnetograms with an average value of 50 G. This is approximately 20 G lower than the standard deviation of the magnetic field strength of the original unsigned magnetograms. The RMS differences decrease on overage by 9 G if the sunspots are masked out.

\begin{figure}
	\centering
	\includegraphics[width=1\linewidth]{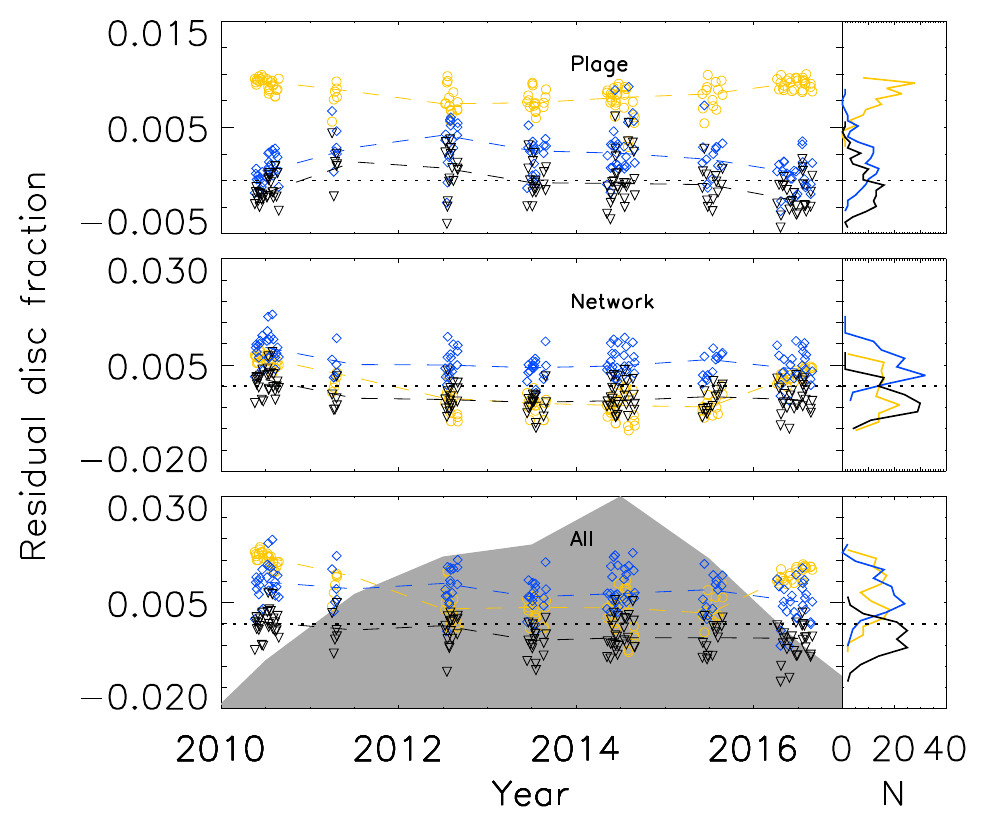}
	\caption{{\it Left:} Difference between the disc fraction of magnetic features derived from SDO/HMI observations with reduced spatial resolution and Rome/PSPT (black downward triangles), SDO/HMI with original spatial resolution (yellow circles), and reconstructed magnetograms with PFL$^*$ (blue rhombuses). Each of the upper two panels corresponds to a different feature (as marked in each panel) identified with Method 1, while the bottom panel is for all features together. The dashed lines connect annual median values, while the dotted horizontal lines are for a difference of 0. The shaded grey surface in the lower panel shows the plage areas determined with Method 1 from the Rome/PSPT images. The areas were scaled to match the range of the plot and are included to indicate the level of solar activity.  {\it Right:} Distribution functions of the residual disc fractions.}
	\label{fig:res_disc_fractions_plage_actnet}
\end{figure}

We also evaluated how well the regions with strong magnetic fields in the reconstructed unsigned magnetograms correspond to magnetic regions and network in the original magnetograms. For this, we derived the disc fractions covered by features with Method 1 (i.e. applying constant thresholds on contrast in the Rome/PSPT and on $|B_{\mathrm{LOS}}|/\mu$ in the original SDO/HMI and reconstructed images).
The residual disc fractions between the ones derived from the degraded unsigned magnetograms and the reconstructed ones with PFL$^*$ are shown in Fig. \ref{fig:res_disc_fractions_plage_actnet}. 
We also show separately the disc fractions derived from the original-size SDO/HMI magnetograms. When doing so we used the same segmentation parameters for all cases. 
For all feature classes the difference of the disc fractions derived from degraded SDO/HMI magnetograms and Rome/PSPT are on average 0.3\% and are always below 1.3\%.
We notice that the area of the features in the degraded SDO/HMI magnetograms increase in disc fraction 
on average by 0.8\% (and up to 1.8\%) compared to the original-sized magnetograms.
The differences between the degraded magnetograms and the reconstructed ones with PFL$^*$ are on average 0.8\% and are always below 2.0\%. We noticed that the reconstructed magnetograms exhibit higher disc fractions for network by $\sim$1\%.

\begin{figure}
	\centering
	\includegraphics[width=1\linewidth]{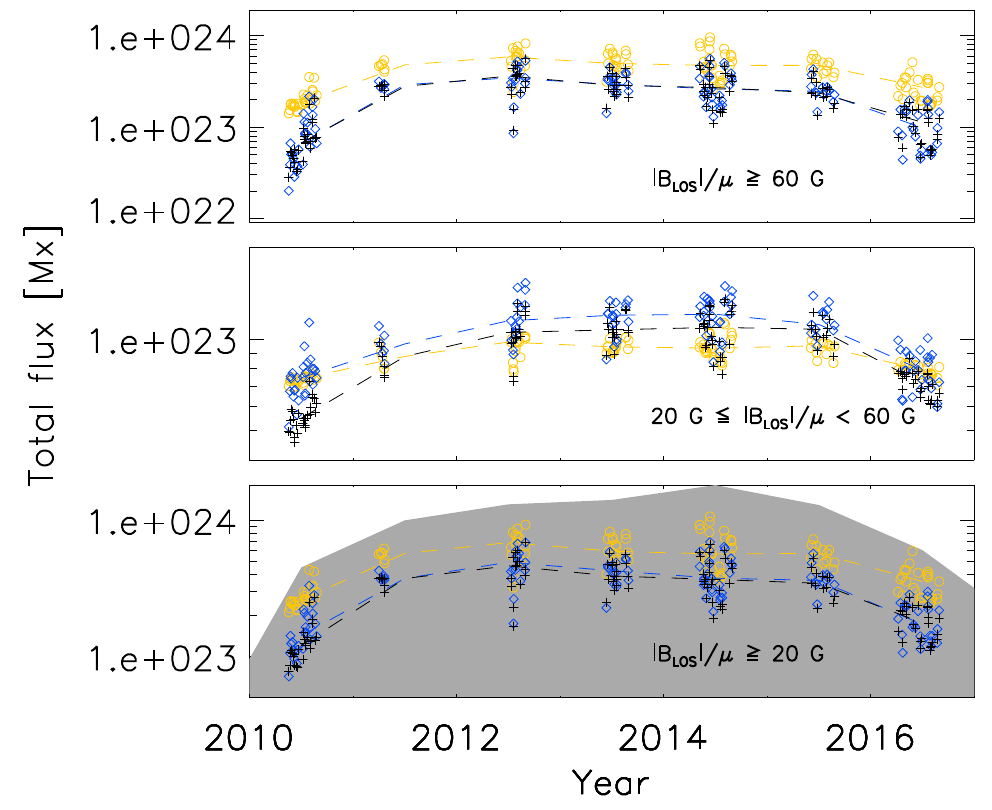}
	\caption{Total unsigned magnetic flux in Mx of AR derived from the magnetograms (yellow circles for the original and black plus signs for the reduced spatial resolution ones) and from the unsigned magnetograms reconstructed from Ca~II~K observations with PFL$^*$ (blue rhombuses). Each of the upper two panels corresponds to a different type of features (as listed in the panels) identified with Method 1, while the bottom panel is for all features together. The dashed lines connect the annual median values. The shaded surface in the lower panel is as in Fig. \ref{fig:res_disc_fractions_plage_actnet}. }
	\label{fig:totflux}
\end{figure}

Finally, we have calculated the total unsigned magnetic flux from the reconstructed and the original unsigned magnetograms. The results are plotted in Fig. \ref{fig:totflux} for the same $|B_{\mathrm{LOS}}|/\mu$ ranges as in Fig. \ref{fig:res_disc_fractions_plage_actnet}.
The day by day correlation coefficient between the total flux in the original and the reconstructed magnetograms with the PFL$^*$ fit is 0.98 for all bright features and is similar for plage and network. 
We noticed that the slightly higher network disc fractions result in an almost constant offset in the total unsigned magnetic flux of the network component. 
The total unsigned magnetic flux in the degraded magnetograms is reduced compared to that from the original-sized magnetograms due to the smoothing applied on the magnetograms to match the Rome/PSPT resolution (see Sec. \ref{sec:datamethods}).


\section{Summary and conclusions}
\label{sec:bvscaksummary}
We have analysed the relationship between the excess Ca~II~K emission and the magnetic field strength. For this, we used 131 sets of co-aligned near-co-temporal SDO/HMI magnetogram and continuum observations and Rome/PSPT filtergrams taken in the core of the Ca~II~K line and in the red continuum. We confirm the existence of a consistent relation between the excess Ca~II~K emission and the magnetic field strength. 
We fit the relation between the Ca~II~K intensity and the vertical component of the magnetic field ($|B_{\mathrm{LOS}}|/\mu$) with a power law function of the logarithm of $|B_{\mathrm{LOS}}|/\mu$ with an offset, and test it against a power law function and a logarithmic function of $|B_{\mathrm{LOS}}|/\mu$ that have been presented in the literature. 
The parameters we derived for the power law function are consistent with those from previous studies. The results for a power law function of $|B_{\mathrm{LOS}}|/\mu$ are also very similar to those derived with a power law function of the logarithm of $|B_{\mathrm{LOS}}|/\mu$. 
The logarithmic function, recently employed in the analysis of high resolution Sunrise data in the Ca II H line, is found to be not representative of bright features in the full-disc Ca~II~K images analysed in our study. 
We note that in previous studies the data were binned in terms of $|B_{\mathrm{LOS}}|/\mu$ before performing the fit. However, results obtained by such fits suffer from 
attenuation bias due to errors in the independent variable, which are not taken into account. For that reason we decided to bin the data both in $|B_{\mathrm{LOS}}|/\mu$ and Ca II K contrast values and perform the fits on the bisector of the two binned curves.

The observations analysed here greatly extend the sample of studied data with respect to previous works.
In particular, we examined a greater amount and in many ways higher-quality data than has been done before for such studies. The data are spanning half a solar cycle, and for this time-scale we report no significant variation with time of the obtained power-law exponents.
Moreover, we find no variation of the exponents over the disc positions between $\mu=1$ and $\mu=0.14$. Finally, the numerical values of the exponents remain nearly the same if stray-light is taken into account. 
We found no significant differences between results derived from images corrected with the methods by \cite{yeo_point_2014} and by \cite{criscuoli_photometric_2017}. 
Having studied this relation for almost the entire disc, up to $\mu=0.14$ or $0.99R$, makes this analysis more applicable to stellar studies than most earlier investigations.

The fact that the exponents are independent of time and $\mu$ suggests that maps of the unsigned LOS magnetic field can be reconstructed from Ca~II~K observations with merely the knowledge of the exponent derived here. We tested the reconstruction of unsigned magnetograms from available Ca II K observations and compared our results to co-temporal directly measured magnetograms. 
The total magnetic flux calculated for the series of the original and reconstructed magnetograms agrees well with the correlation factor of 0.98. This means that historical Ca II K spectroheliograms, when properly processed and calibrated \citep[e.g.][]{chatzistergos_exploiting_2016,chatzistergos_ca_2018,chatzistergos_historical_2019,chatzistergos_analysis_2019}, can be used to extend the series of magnetograms throughout the whole 20th century. This approach suffers from the limitation that it does not allow the polarity of the magnetic field to be recovered. However, this is not a problem for a number of studies and applications, e.g. for irradiance reconstructions, where the models do not require the polarity of the bright features.
Besides, if other data are also used, for instance sunspot measurements, it might be possible to recover the polarity of the ARs as well.

\begin{acknowledgements}
We are grateful to Yang Liu, Aimee Norton, and Phil Scherrer for the helpful and fruitful discussions and for providing the stray-light corrected SDO/HMI data. 
T.C. acknowledges a postgraduate fellowship of the International Max Planck Research School on Physical Processes in the Solar System and Beyond.
I.E. acknowledges support by grants PRIN-INAF-2014 and PRIN/MIUR 2012P2HRCR "Il Sole attivo" and FP7 SOLID. The work was also partly supported by COST Action ES1005 "TOSCA", and by the BK21 plus program through the National Research Foundation (NRF) funded by the Ministry of Education of Korea. This research has made use of NASA's Astrophysics Data System.
\end{acknowledgements}

\bibliographystyle{aa}
\bibliography{_biblio1}

\appendix
\section{Spatial agreement between magnetograms and Ca II K images}
\label{sec:spatialagreement}
\begin{figure}[t!]
	\centering
	\includegraphics[width=1\linewidth]{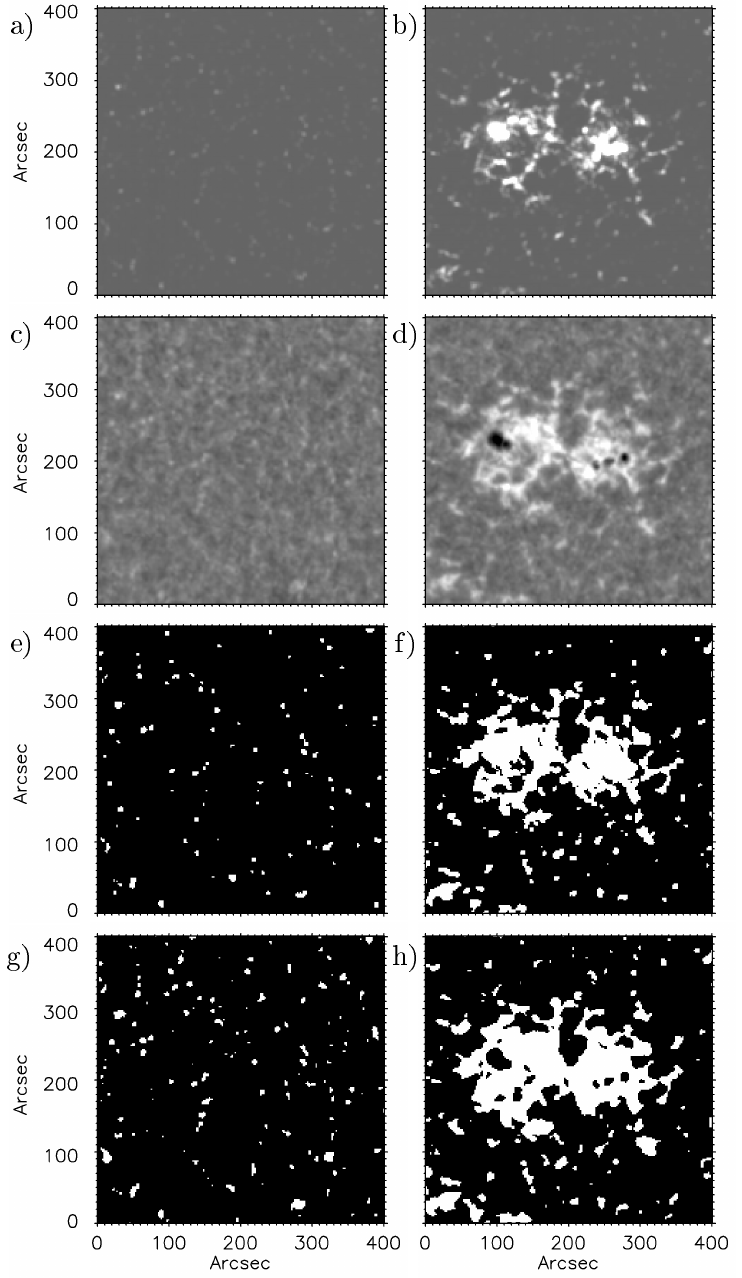}
	\caption{Magnified 400$\times$400 arcsec$^2$ sub-arrays of the images shown in Fig. \ref{fig:psptcaexamples} for a network (left) and a plage region (right). From top to bottom: (a), b)) SDO/HMI unsigned (and spatially degraded) magnetogram; (c), d)) Rome/PSPT Ca~II~K; the corresponding segmentation masks derived with Method 1, i.e. constant thresholds, from (e), f)) the magnetograms and (g), h)) the Ca~II~K images. The magnetograms are saturated in the range [-300,300] G (the negative value was chosen merely to improve visibility of the pixels), and Ca~II~K observations in the range [0.5,1.6] (the QS has the average value of 1).}
	\label{fig:networkzoom}
\end{figure}

Figure \ref{fig:networkzoom} displays examples of close-ups of one active and one quiet region from the observations shown in Fig. \ref{fig:psptcaexamples} to illustrate the good spatial agreement between the SDO/HMI and Rome/PSPT images.
Figure \ref{fig:networkzoom} e)--h) displays the corresponding masks of plage and network combined for the close-ups shown in Fig. \ref{fig:networkzoom} a)--d) derived with method 1 (see Sect. \ref{sec:segmentation}). Figure \ref{fig:networkzoom} illustrates the well-known fact that the bright features  in the Ca~II~K images belong to magnetic regions and network in the magnetograms.
The ARs appear slightly smaller and show smaller-scale features in the
magnetograms than in the Ca~II~K data. This can occur for a variety of reasons. The flux tubes comprising ARs expand with height in the solar atmosphere, therefore ARs are expected to be more extended in Ca II K images. Furthermore, if the flux tubes are inclined then they can appear more broadened in the Ca~II~K data too. Other possible reasons include lower spatial resolution and seeing effects due to Earth's atmosphere that smear the features in the Ca~II~K observations. 
Some contribution will be provided by cancellation of magnetograph signal between opposite polarities within the same resolution element \citep[see e.g.][for hidden opposite polarities at SDO/HMI resolution that appear at the higher resolution of Sunrise observations]{chitta_solar_2017}. 
However, these effects should be minimised after the spatial degradation we applied to the magnetograms. Finally, the choice of the segmentation thresholds has an effect as well, if they are not consistent between the magnetograms and Ca~II~K images. We evaluated a variety of threshold combinations, but we were unable to match better the AR areas in the two observations without introducing even smaller scale features in the magnetograms. Therefore, we assumed that the differences are to a significant extent due to the expansion of the flux tubes, in particular by the fibrils spreading out at the edges of ARs, as found by, e.g., \cite{pietarila_bright_2009}.

\section{Reconstructed magnetograms with different functions}
\label{sec:appendix_comparison}
\begin{figure*}[t!]
	\centering
	\includegraphics[width=1\linewidth]{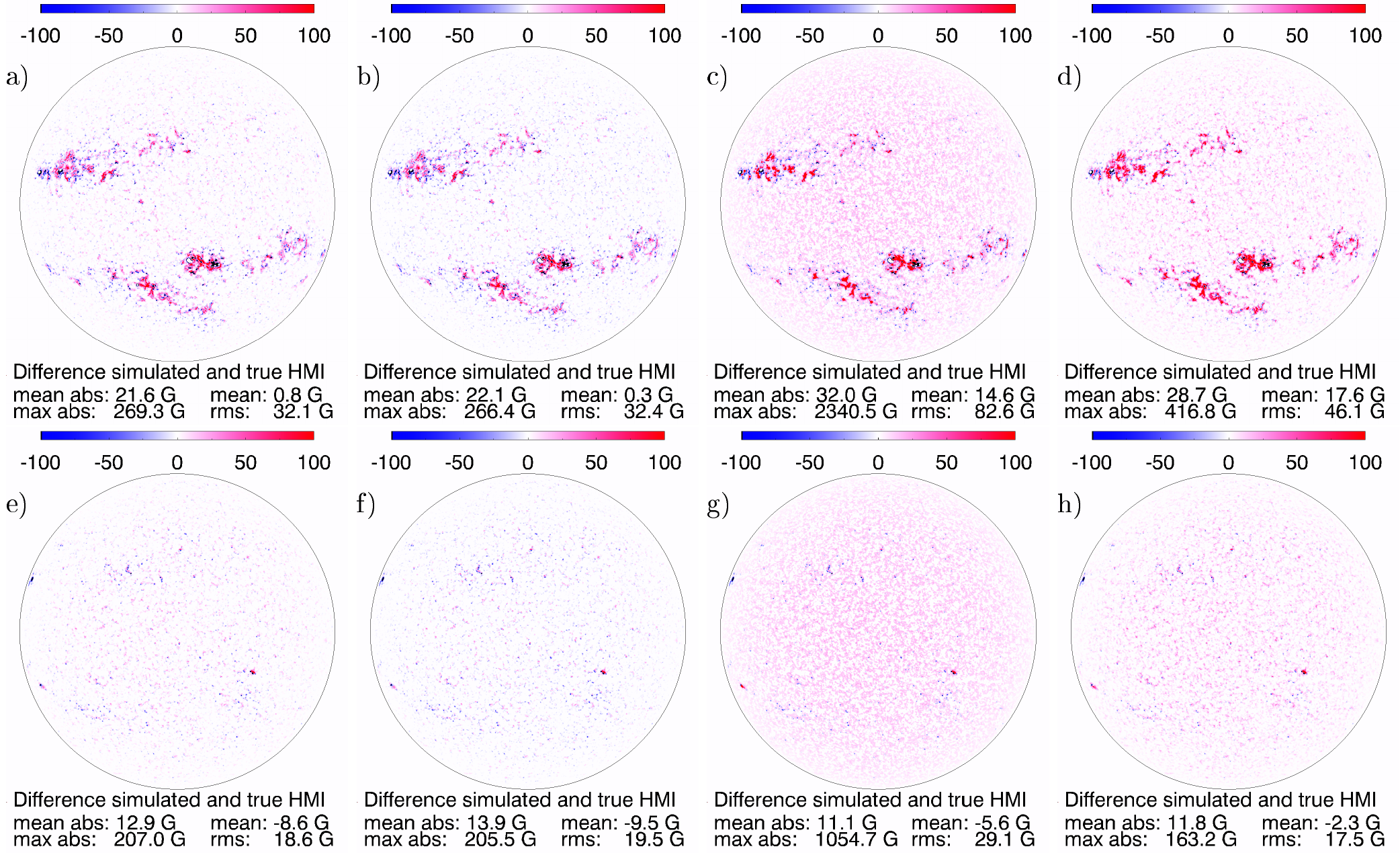}
	\caption{Difference between the reconstructed unsigned magnetogram (simulated) from Ca~II~K data and original (true) SDO/HMI unsigned magnetogram taken on 01/04/2011 (top) and 07/06/2010 (bottom). The reconstruction was done with the parameters derived in Sect. \ref{sec:scatterpots} for PF$^*$ (a), e)), PFL$^*$ (b), f)), LFL$^*$ (c), g)), and PF (d), h)). The RMS, mean, mean absolute, and maximum absolute differences are listed under each panel. The colour bars show the ranges of $B$ in G.}
	\label{fig:20000821nsb2}
\end{figure*}
Here we use the parameters derived for PF, PF$^*$, and LFL$^*$ to reconstruct unsigned magnetograms and compare the results with those derived with PFL$^*$.
Figure \ref{fig:20000821nsb2} shows the pixel by pixel absolute differences between the reconstructed and the original magnetograms by using PF$^*$ (panels a), e)), PFL$^*$ (panels b), f)), LFL$^*$ (panels c), g)), and PF (panels d), h)).

Comparing the errors between the reconstructed and the original magnetograms 
we got similar uncertainties for both PF$^*$ and PFL$^*$. In particular we found RMS differences of $\simeq30$ G and $\simeq20$ G for the active and quiet day, respectively, for both PF$^*$ and PFL$^*$. We discern no significant difference between these two reconstructed magnetograms, although a careful comparison reveals many differences at small scales. The differences for the quiet day show that we slightly underestimated the weak fields. 
The differences for the LFL reach up to 2500 G in plage regions. These high errors arise due to the large pixel-to-pixel scatter in the relationship between Ca II K contrast and $|B_{\mathrm{LOS}}|/\mu$. Consequently there are numerous very bright pixels in the Ca~II~K observations that would correspond to very strong fields in this case, as the fitted curve increases very slowly. This problem is somewhat more acute for reconstructions that use the PF and PFL relationships (i.e. those derived from a fit to data binned in $|B_{\mathrm{LOS}}|/\mu$). We also show the differences for PF, which has been commonly used in the literature. In this case, the errors are slightly higher than for PF$^*$ or PFL$^*$ for times with high activity. 

\begin{figure}
	\centering
	\includegraphics[width=1\linewidth]{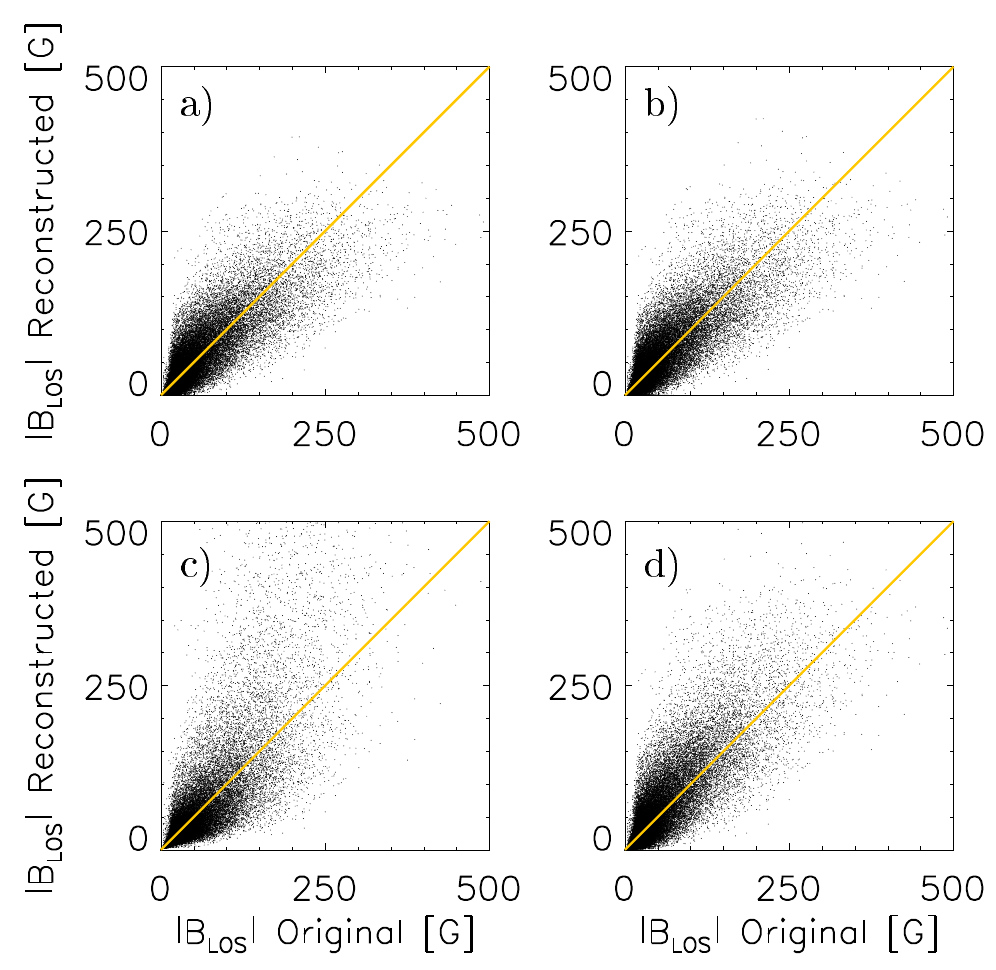}
	\caption{Scatter plots between original (degraded) magnetograms and those reconstructed from the Ca~II~K image taken on 01/04/2011 using the average parameters for PF$^*$ (a)), PFL$^*$ (b)), LFL$^*$ (c)), and PF (d)). The yellow line has a slope of unity. The axes are shown in the range from the original magnetogram.}
	\label{fig:scatterplots_originalreconstructedmagnetograms_appendix}
\end{figure}

\begin{figure}
	\centering
	\includegraphics[width=1\linewidth]{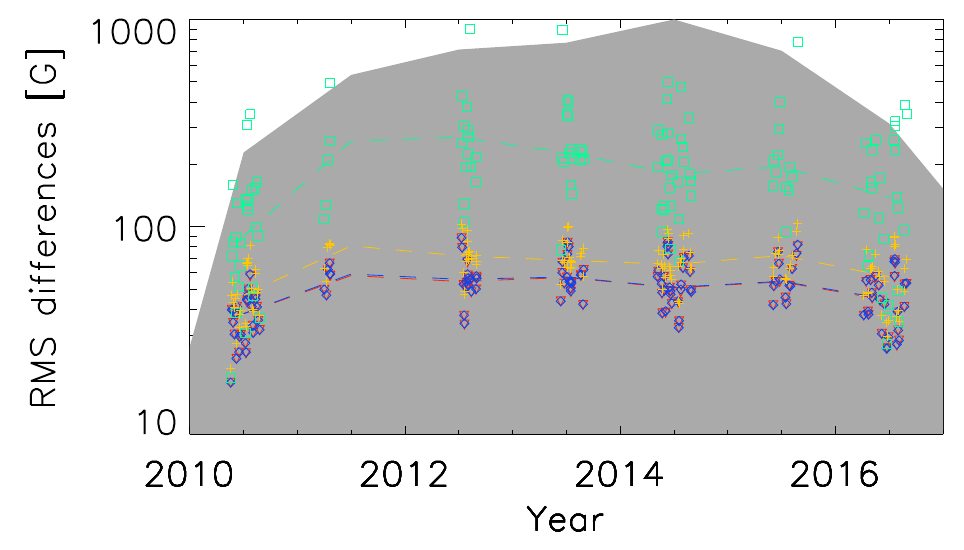}
	\caption{RMS pixel by pixel differences in G between the original magnetograms and the reconstructed ones using the parameters (listed in Table \ref{tab:functionparameters}) derived from the PF$^*$ (red), PFL$^*$ (blue), LFL$^*$ (green), and PF (yellow) fits. The dashed lines connect annual median values. The shaded surface is as in Fig. \ref{fig:res_disc_fractions_plage_actnet}. }
	\label{fig:differences25070_diffits}
\end{figure}

Figure \ref{fig:scatterplots_originalreconstructedmagnetograms_appendix} shows scatter plots between the four reconstructed magnetograms and the original one for the observation taken on 01/04/2011 (the active day shown in Fig. \ref{fig:20000821nsb2}). The unsigned magnetograms reconstructed with PF$^*$ and PFL$^*$ show the best correspondence, while the ones with LFL$^*$ and PF tend to overestimate the magnetic field. 
Figure \ref{fig:differences25070_diffits} shows the pixel by pixel RMS differences  between the original and the reconstructed unsigned magnetograms obtained using the derived best fit parameters of the three functions we tested, without masking the surroundings of the sunspots this time.
Figure \ref{fig:differences25070_diffits} reveals that the RMS differences remain less than 88 G for all 131 unsigned magnetograms reconstructed with the PF$^*$ and PFL$^*$, but reach 100 G for PF and 7500 G for LFL$^*$.

\begin{figure}
	\centering
	\includegraphics[width=1\linewidth]{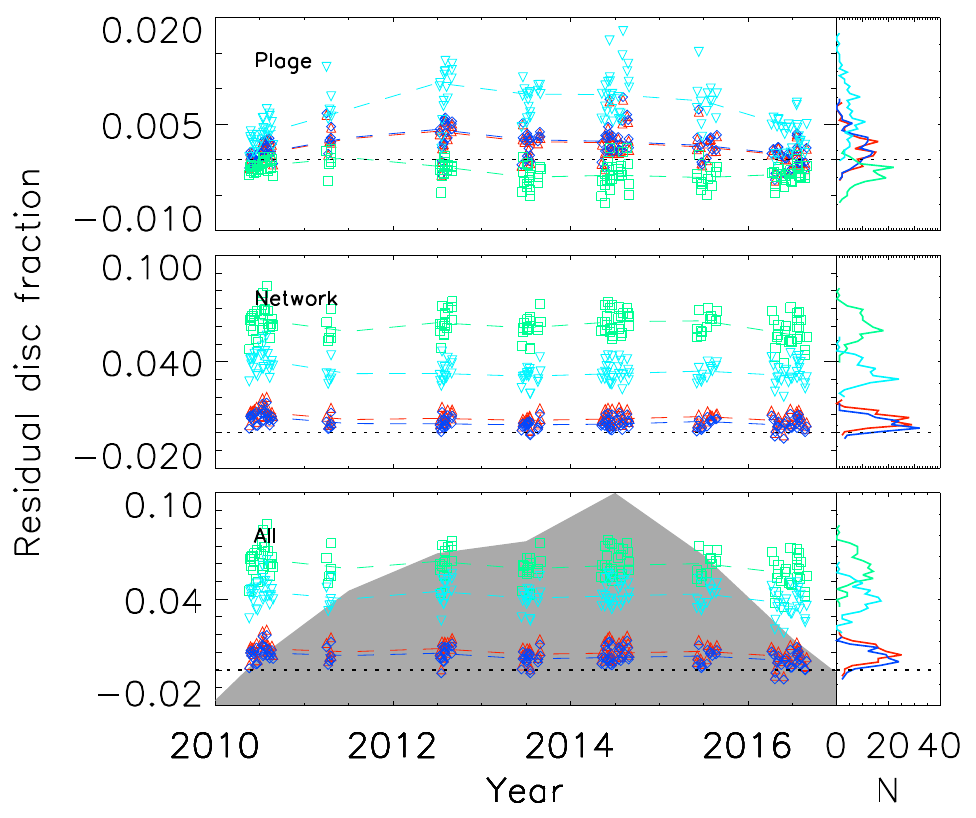}
	\caption{{\it Left:} Difference between the disc fraction of magnetic features derived from SDO/HMI observations with reduced spatial resolution and reconstructed magnetograms with PF$^*$ (red upward triangles), PFL$^*$ (blue rhombuses), LFL$^*$ (green squares), and PF (light blue downward triangles). Each of the upper two panels corresponds to a different feature (as marked in each panel) identified with Method 1, while the bottom panel is for all features together. The dashed lines connect annual median values, while the dotted horizontal lines are for a difference of 0. The shaded surface in the lower panel is as in Fig. \ref{fig:res_disc_fractions_plage_actnet}.  {\it Right:} Distribution functions of the residual disc fractions.}
	\label{fig:res_disc_fractions_plage_actnet_diffits}
\end{figure}

Figure \ref{fig:res_disc_fractions_plage_actnet_diffits} shows the residual disc fractions between the ones derived from the degraded unsigned magnetograms and the reconstructed ones with the PF$^*$, PFL$^*$, LFL$^*$, and PF fits. 
The results for PF$^*$ follow very closely those for PFL$^*$, though giving minutely (on average by 0.3\%) higher differences.
The differences between the degraded magnetograms and the reconstructed ones with PF$^*$ fits are on average 1.0\% and are always below 2.3\%. 
The disc fractions in the magnetograms reconstructed with LFL$^*$ are on average 6\% higher than in the original magnetograms when all features are considered, however the difference remains less than 0.1\% when only the plage regions are considered.
The errors in the disc fractions slightly increase when the magnetograms are reconstructed with PF, being $\sim$4\% for all features. 

\begin{figure}
	\centering
	\includegraphics[width=1\linewidth]{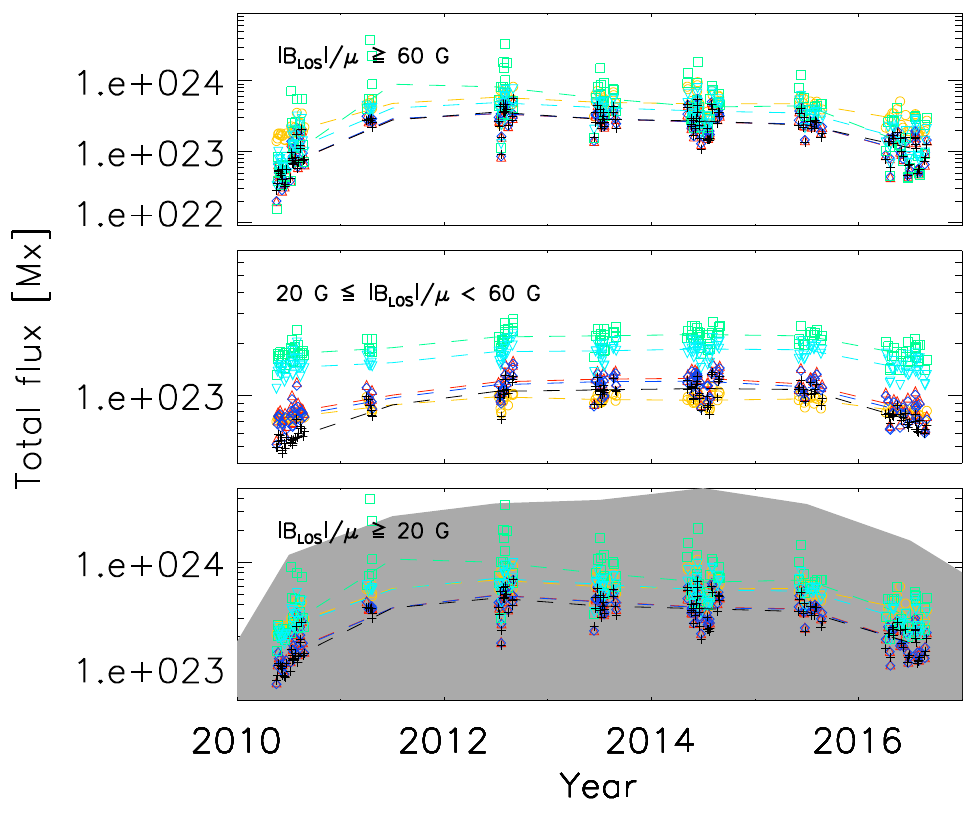}
	\caption{Total unsigned magnetic flux in Mx of AR derived from the magnetograms (yellow circles for the original and black plus signs for the reduced spatial resolution ones) and from the unsigned magnetograms reconstructed from Ca~II~K observations with PF$^*$ (red upward triangles), PFL$^*$ (blue rhombuses), LFL$^*$ (green squares), and PF (light blue downward triangles). Each of the upper two panels corresponds to a different type of features (as listed in the panels) identified with Method 1, while the bottom panel is for all features together. The dashed lines connect the annual median values. The shaded surface in the lower panel is as in Fig. \ref{fig:res_disc_fractions_plage_actnet}. }
	\label{fig:totflux_diffits}
\end{figure}

The total unsigned magnetic flux is plotted in Fig. \ref{fig:totflux_diffits} for the same $|B_{\mathrm{LOS}}|/\mu$ ranges as in Fig. \ref{fig:res_disc_fractions_plage_actnet_diffits}.
The day by day correlation coefficient between the total flux in the original and the reconstructed magnetograms with both PF$^*$ and PFL$^*$ fits is 0.98 for all bright features and is similar for plage and network. The differences between the results for PF$^*$ and PFL$^*$ are minute, with the latter giving slightly higher values. The total flux derived from the reconstructed unsigned magnetograms with PF and LFL$^*$ give consistently higher values.

\clearpage


\end{document}